\documentclass[aps,prd,prabib,showpacs,nofootinbib] {revtex4}
\usepackage{graphicx} \usepackage{amsmath} \usepackage{amssymb}
\usepackage{amsfonts} \usepackage{bm}
\usepackage{array}
\usepackage{siunitx}
\usepackage[singlelinecheck=false]{caption}
\usepackage{ytableau}
\usepackage{multirow}
\newcommand*{\Comb}[2]{{}^{#1}C_{#2}}

\begin{document}

\newcommand{\be}{\begin{equation}} \newcommand{\ee}{\end{equation}}
\newcommand{\bea}{\begin{eqnarray}}\newcommand{\eea}{\end{eqnarray}}

\title{Singular eigenstates  in the even(odd) length Heisenberg spin chain}

\author{Pulak Ranjan Giri} \email{pulakgiri@gmail.com}

\author{Tetsuo Deguchi} \email{deguchi@phys.ocha.ac.jp}

\affiliation{ Department of Physics, Graduate School of Humanities and Sciences, Ochanomizu University, Ohtsuka 2-1-1, Bunkyo-ku, Tokyo, 112-8610, Japan}

\begin{abstract}
We study the implications of the regularization for the singular solutions on  the   even(odd) length  spin-$1/2$ XXX  chains in some specific  down-spin sectors.  In particular,  the analytic expressions  of the Bethe eigenstates for three down-spin sector   have been obtained along with their    numerical  forms   in some fixed length chains. 
For an  even-length  chain if  the  singular solutions  $\{\lambda_\alpha\}$  are invariant under the sign changes  of their rapidities    $\{\lambda_\alpha\}=\{-\lambda_\alpha\} $,  then the Bethe ansatz equations are  reduced  to a system  of  $(M-2)/2 ((M-3)/2)$ equations   in    an even (odd)  down-spin sector.
For an  odd  $N$ length   chain  in the  three down-spin sector,  it has been analytically  shown that  there exist  singular  solutions in  any finite  length  of the spin chain of the form  $N= 3\left(2k+1\right)$ with  $k=1, 2, 3, \cdots$.   It is also shown that  there exist no  singular solutions in  the  four down-spin sector for some  odd-length     spin-$1/2$  XXX  chains. 
\end{abstract}


\date{\today}

\maketitle

\section{Introduction}
More than eight decades ago, Bethe solved  \cite{bethe}  the  spin-$1/2$  isotropic Heisenberg   chain, i.e.  the spin-$1/2$  XXX  chain,   by a method, known as  the  Bethe ansatz.   In the  algebraic Bethe ansatz 
\cite{faddeev,koma,korepin,faddeev1,mart,bei}, the  eigenvalues and the  eigenstates    are expressed  in terms of   the  rapidities   $\lambda_\alpha$, known as the Bethe roots.   These $\lambda_\alpha$ are the solutions of  the Bethe ansatz equations, which are  a set of polynomial  equations, emerge  as  conditions   for the  eigenvalue  equation of the transfer matrix of the  spin-$1/2$ XXX  chain.   Numerical methods, such as, the  Newton-Raphson,  homotopy continuations and   iterations  are usually    deployed  to solve the Bethe ansatz equations.    The distinct and self-conjugate  solutions  \cite{vlad2} of the Bethe ansatz equations  produce   the  Bethe eigenstates  of the  spin-$1/2$  XXX chain which are of  highest weight.  For the higher spin chains, however,  there are   repeated rapidities \cite{vlad1} in  some  solutions, which produce  the  Bethe  eigenstates.   The  complex  solutions   present more  challenges numerically as opposed to the real solutions, which are easier to evaluate. 

Nonetheless, there has been  growing interest  in the  solutions of the   spin-$1/2$  XXX chain   in recent years \cite{fuji,hag,nepo1,giri}.   Although  making use of   the  {\it string hypothesis}  \cite{taka}  one can estimate the total number of  Bethe eigenstates, its certain assumptions  do  not always  hold for any  given  finite length spin chain.  For example, as the length of the chain increases,   some of the two string solutions deform back  to form two real distinct  rapidities  \cite{essler,isler,ila} and some of the  two strings have much larger  rapidities \cite{vlad} for  very large length spin  chains,  which are  a violation of  the  {\it string  hypothesis}. 
It is therefore necessary to look into   the  detailed  analysis  of the  Bethe ansatz solutions.   Moreover effects of the complex solutions on quantities such as  the  correlation  functions \cite{maillet,maillet1}, form factors and   fidelity are also  important, while we  need complete knowledge of the complex  solutions beforehand in order to investigate them explicitly.   It is also worth to mention  that  some  types of solutions of the Bethe ansatz equations in the anisotropic Heisenberg spin chains  are studied in \cite{woy,bab,fab1,fab2,td}.

The  sets of rapidities associated with the spectrum of  the   spin-$1/2$ XXX  chain   are  of two classes. One is  regular solutions, for which  both the  Bethe eigenstates  and the eigenvalues are   finite and well-defined.    The other is the 
singular sets of  rapidities  \cite{sid,noh}, which have  one pair of rapidities  of the form $\{\lambda_1=\frac{i}{2}, \lambda_2=-\frac{i}{2}\}$.  As the name suggests, the  Bethe eigenstates  and the eigenvalues are  ill-defined  because of the  pair 
$\{\lambda_1=\frac{i}{2}, \lambda_2=-\frac{i}{2}\}$. If one straightforwardly plugs  the singular solutions into the  formula for the  Bethe eigenstates  in the algebraic Bethe ansatz method  or into  the eigenvalues, then the states vanish  and  the eigenvalues diverge.   Singular solutions, nevertheless, are an essential part of the  spectrum,  because, without them   the solutions are not complete.     It is therefore imperative to  devise  a  regularization scheme   \cite{bei,vlad1,nepoh,goe,aru,nepot,nepo1,nepo,kirillov2}
 to make the singular solutions viable such that both the  eigenvectors and the eigenvalues become finite and well-defined.  Recently,  a  detailed  investigation  is  carried out  by Nepomechie and Wang   \cite{nepo} and extended to higher spin chains \cite{nepoh}, where the authors   first solve the pole free form of the Bethe ansatz equations for the singular solutions and then introduce  the regularization scheme to   obtain a consistency condition, which is  satisfied    only by  the physical singular solutions
(i.e.   the solutions which do  produce  the Bethe eigenstates and   their   corresponding eigenvalues).    We note that in the standard approach for  solving the algebraic Bethe ansatz  there is an implicit assumption that   no Bethe roots contain rapidities of the form  $\pm \frac{i}{2}$.  As mentioned above,   the presence of   $\pm \frac{i}{2}$ reduce  the Bethe eigenstates   to null states,  making the eigenvalue equation trivial.

The purpose of this paper is to  study the implications of the already developed regularization scheme on the   even(odd) length spin chains in some specific  down-spin sectors. 
For an even length  spin-$1/2$ XXX  chain,    the singular solutions  which  are invariant under  the  change of   sign of each of the rapidities,  i.e.   $\{\lambda_\alpha\}= \{-\lambda_\alpha\}$, simplify     the Bethe ansatz equations significantly   such that they can be handled   easily in the numerical  process.    For example, 
in  our previous work  \cite{giri}  on  non self-conjugate strings, singular strings and rigged configurations \cite{kirillov1,kirillov3,kirillov4,lule,kirillov,tho}  of  the  spin-$1/2$ XXX chain, it helped us obtain the singular solutions  in  specific cases  easily.  We analytically  show  that    the singular solutions  $\{\lambda_1= \frac{i}{2}, \lambda_2= -\frac{i}{2}, \lambda_3= \pm \frac{\sqrt{3}}{2}\}$  are present  for  any odd-length  chain of  the form of  $N= 3\left(2k+1\right)$ with  $k=1, 2, 3, \cdots$.  The repetition of these singular solutions  with  such a periodicity of $6$ in  $N$  has  already  been   confirmed numerically   in \cite{nepo1}  for some  values   of  the length of the spin  chain.    Analytically explicit expressions of   the Bethe eigenstates  for $M=3$ have been obtained for even and odd-length spin chains and  the numerical forms of   these states are  also obtained for some fixed lengths.  A   graphical method  is provided  to search for  any  singular solution  present, if at all,  for  the   $M=4$  sector   in   some finite  odd-length spin chains.

We organize this paper in the following fashion:  In the next section,  we   briefly discuss   the  algebraic Bethe ansatz method for the   spin-$1/2$ XXX  chain, which sets the basis for the subsequent  sections. In section III,   we review   the  regularization  for the singular solutions, which  has been      studied    recently  in  ref. \cite{nepo}. 
In section IV  we show for the even-length  spin-$1/2$ XXX  chain that  the  Bethe ansatz equations for the singular solutions 
such that they  are  symmetrically distributed  in the complex plane of rapidities,  i.e.   $\{\lambda_\alpha\}=\{-\lambda_\alpha\} $, can be  written in  a significantly reduced form.   
The explicit  expression   of   the three down-spin singular Bethe eigenstate  for even-length chains   has   been obtained  and    a   derivation of the formulae for  the  Bethe eigenstate with  two down-spins  and that of three down-spins  in   the even-length chain have    been provided in {\bf Appendix \ref{app1}} and   {\bf Appendix \ref{app2}},  respectively. 
In section V  it  is   analytically shown that  there exist singular solutions of the form   $\{\lambda_1= \frac{i}{2}, \lambda_2= -\frac{i}{2}, \lambda_3= \pm \frac{\sqrt{3}}{2}\}$  in any odd-length chain  of the form  $N= 3\left(2k+1\right)$ with  $k=1, 2, 3, \cdots$. It  is  shown  in  {\bf Appendix  \ref{app3}}.  The corresponding Bethe eigenstates are derived in {\bf Appendix \ref{app21}}.  A graphical method  is also  suggested  for the  odd $N$ cases   to search for any possible singular solutions in  the $M=4$ down-spin sector  and we show that  for $N=15, M=4$ there is no singular solutions. 
Finally we conclude  in section VI.  
%------------------------------------------------------------------------------------------------------------------------------------------------------------------------------------------

\section{Algebraic Bethe Ansatz}
The  spin-$1/2$  XXX chain on a one-dimensional periodic lattice  of  length $N$  is given by the Hamiltonian   
\begin{eqnarray}\label{ham}
H= J\sum_{i=1}^{N}\left(S^x_iS^x_{i+1}+S^y_iS^y_{i+1}+S^z_iS^z_{i+1}- \frac{1}{4}\right)\,,
\end{eqnarray}
where  $J$ is the coupling constant and $S_i^j (j=x,y,z)$  is  the  spin-$1/2$ operator at the  $i$-th lattice site  and in $j$-direction.   The eigenstates and eigenvalues of this Hamiltonian can be obtained in the algebraic Bethe ansatz formulation in the following way. Let us consider the Lax operator  as
\begin{eqnarray}
L_\gamma(\lambda)=\left( \begin{array}{cc}
\lambda-iS^z_\gamma & -iS^-_\gamma  \\
-iS^+_\gamma & \lambda+iS^z_\gamma  \end{array} \right)\,,
\end{eqnarray}
where $S^\pm_\gamma= S^x_\gamma \pm iS^y_\gamma$ and  each element  of $L_\gamma(\lambda)$ is  a matrix of dimension  $2^N \times 2^N$, which  acts nontrivially  on the $\gamma$-th lattice site.  The monodromy matrix, $T(\lambda)$, is then given by the  direct product of  the Lax matrices  at each site 
\begin{eqnarray}\label{monodromy}
T(\lambda)=L_N(\lambda)L_{N-1}(\lambda) \cdots L_1(\lambda)=\left( \begin{array}{cc}
A(\lambda)& B(\lambda)  \\
C(\lambda) & D(\lambda) \end{array} \right)\,.
\end{eqnarray}
The Hamiltonian  (\ref{ham}) can be obtained from the transfer matrix
\begin{eqnarray}\label{tmatrix}
t(\lambda)=  A(\lambda)+D(\lambda)\,,
\end{eqnarray} 
by taking its  logarithm  at  $\lambda= -\frac{i}{2}$ as 
\begin{eqnarray}
H=  \frac{J}{2}\left(-i\left[\frac{d}{d\lambda}\log t(\lambda)\right]_{\lambda=-\frac{i}{2}}-N\right)\,.
\end{eqnarray} 

In terms of the  rapidities  $\lambda_\alpha$, the  Bethe  state  in    the  $M$ down-spin sector  is  expressed as 
\begin{eqnarray}\label{vec0}
|\lambda_1, \lambda_2,\cdots,\lambda_M\rangle= \prod_{\alpha=1}^MB(\lambda_\alpha)|\Omega\rangle\,,
\end{eqnarray}
where  $|\Omega\rangle$  is the reference eigenstate with all  spins  up and  $B(\lambda_\alpha)$   is  an element of  the monodromy matrix $T(\lambda_\alpha)$ obtained from  eq. (\ref{monodromy}). 
The Bethe  state (\ref{vec0}) can  explicitly be written  as  \cite{deguchi1}
\begin{eqnarray}\label{vec1}\nonumber
\prod_{\alpha=1}^MB(\lambda_\alpha)|\Omega\rangle= &&(-i)^M \prod_{j < k}^{M}\frac{\lambda_j-\lambda_k +i}{\lambda_j-\lambda_k} \prod_{j =1}^{M}\frac{(\lambda_j- \frac{i}{2})^N}{\lambda_j+ \frac{i}{2}}\times \\
&&\sum_{1\leq x_1 < x_2 \cdots < x_M \leq N}^{N}\sum_{\mathcal{P}\in S_M}^{M!} \prod_{\mathcal{P}j <\mathcal{P}k}^{M} \left(\frac{\lambda_{\mathcal{P}j}-\lambda_{\mathcal{P}k}-i}{\lambda_{\mathcal{P}j}-\lambda_{\mathcal{P}k}+i}\right)^{H(j-k)} \prod_{j=1}^M\left(\frac{\lambda_{\mathcal{P}j}+ \frac{i}{2}}{\lambda_{\mathcal{P}j}- \frac{i}{2}}\right)^{x_j}\prod_{j=1}^M S^-_{x_j}|\Omega \rangle\,,
\end{eqnarray}
where $\mathcal{P}$ are elements of   the permutation  group $S_M $  of  $M$ numbers and $H(x)$ is the  Heaviside step function  $H(x) =1$ for $x >0$ and  $H(x) =0$ for $x \leq 0$.   

The action of the transfer matrix   (\ref{tmatrix}) on  the   Bethe state (\ref{vec0})   is given by 
\begin{eqnarray}\label{tb}
t(\lambda)\prod_{\alpha=1}^MB(\lambda_\alpha)|\Omega\rangle= \Lambda\left(\lambda, \{\lambda_\alpha\}\right)\prod_{\alpha=1}^MB(\lambda_\alpha)|\Omega\rangle + \sum_{k=1}^{M} 
\Lambda_{k}\left(\lambda, \{\lambda_\alpha\}\right)B(\lambda)\prod_{\alpha \neq k}^MB(\lambda_\alpha)|\Omega\rangle\,,
\end{eqnarray}
where  
\begin{eqnarray}\label{teigen}
\Lambda\left(\lambda, \{\lambda_\alpha\}\right)= \left(\lambda + \frac{i}{2}\right)^N \prod_{\alpha=1}^M\frac{\lambda-\lambda_\alpha-i}{\lambda-\lambda_\alpha} + 
\left(\lambda - \frac{i}{2}\right)^N \prod_{\alpha=1}^M\frac{\lambda-\lambda_\alpha+i}{\lambda-\lambda_\alpha}\,,
\end{eqnarray}
is   the eigenvalue of the transfer matrix   and   the  unwanted terms are 
\begin{eqnarray}\label{tunw}
\Lambda_k\left(\lambda, \{\lambda_\alpha\}\right)= \frac{i}{\lambda-\lambda_k}\left[\left(\lambda_k + \frac{i}{2}\right)^N \prod_{\substack{{\alpha =1} \\{\alpha\neq k}}}^M\frac{\lambda_k-\lambda_\alpha-i}{\lambda_k-\lambda_\alpha} - 
\left(\lambda_k - \frac{i}{2}\right)^N \prod_{\substack{{\alpha =1} \\{\alpha\neq k}}}^M\frac{\lambda_k-\lambda_\alpha+i}{\lambda_k-\lambda_\alpha}\right]\,,~ k= 1, 2, \cdots,  M\,.
\end{eqnarray}
Note that   (\ref{tb}) becomes an eigenvalue equation when the unwanted terms  (\ref{tunw}) vanish, which  give us the well known   Bethe ansatz equations 
\begin{eqnarray}\label{bethe0}
\left(\frac{\lambda_\alpha- \frac{i}{2}}{\lambda_\alpha+ \frac{i}{2}}\right)^{N}= \prod_{\substack{{\beta =1} \\{\beta\neq \alpha}}}^{M}\frac{\lambda_\alpha-\lambda_\beta-i}{\lambda_\alpha-\lambda_\beta+i}\,, ~~~  \alpha=1,2, \cdots, M\,.
\end{eqnarray}
In terms of  solutions $\lambda_\alpha$ of   (\ref{bethe0}),   known as   the Bethe roots, the eigenvalue  of the Hamiltonian  $H$ for  the  $M$ down-spin state  is expressed  as   
\begin{eqnarray}\label{eigen}
E=   \frac{J}{2}\left(-i\left[\frac{d}{d\lambda}\log \Lambda\left(\lambda, \{\lambda_\alpha\}\right)\right]_{\lambda=-\frac{i}{2}}-N\right)= -J\frac{1}{2}\sum_{\alpha=1}^{M}\frac{1}{\left(\lambda_\alpha^2+ \frac{1}{4}\right)}\,.
\end{eqnarray}

To characterize the state in terms of the   Bethe quantum numbers, $\{J_\alpha, \alpha=1,2, \cdots, M\}$,  one  takes  the logarithm of  eq. (\ref{bethe0}) as 
  \begin{eqnarray}\label{logform}
2\arctan(2\lambda_\alpha)= J_\alpha\frac{2\pi}{N} + \frac{2}{N}\sum_{\substack{{\beta =1}\\{\beta\neq\alpha}}}^{M}\arctan(\lambda_\alpha-\lambda_\beta)\,, ~~~  \alpha=1,2, \cdots, M\,, ~~~~\mbox{mod}~2\pi\,.\end{eqnarray}
The  Bethe quantum numbers  take  integral (half integral)  values   if   $N-M$ is odd (even)  respectively.   $J_\alpha$ are  in general repetitive and therefore  are  not much useful  to count the total number of states of  a spin chain.
However, strictly  non-repetitive quantum numbers  can also be obtained.  According to  the {\it string hypothesis},   the rapidities for   the  $M$ down spin   sector are typically arranged in a set of  strings as,
\begin{eqnarray}\label{string}
\lambda_{\alpha a}^{j}= \lambda_{\alpha}^{j} +  \frac{i}{2}\left(j+1-2a\right) +  \Delta_{\alpha a}^{j}\,, ~~~ a=1,2, \cdots, j, ~~\alpha=1,2,.., M_j\,,
\end{eqnarray}
where the string center   $\lambda_\alpha^j$  for a length  $j$-string is real, $\alpha$  represents  the number of  $j$-strings $M_j$  and  the string deviations are given by $\Delta_{\alpha a}^{j}$.  In the limit that the deviations vanish,  $\Delta_{\alpha a}^{j} \to 0$, equations (\ref{logform}) reduce to the     equations 
\begin{eqnarray}\label{betheta}\nonumber
\arctan\frac{2\lambda^j_\alpha}{j} &=& \pi \frac{I^j_\alpha}{N} + \frac{1}{N}\sum_{k=1}^{N_s}\sum_{\beta}^{M_k}\Theta_{jk}\left(\lambda_\alpha^j-\lambda_\beta^k\right)\,,~~~~ \mbox{mod} ~\pi\,,\\
\Theta_{jk}(\lambda) &=& (1-\delta_{jk})\arctan\frac{2\lambda}{|j-k|} + 2\arctan\frac{2\lambda}{|j-k|+2} + \cdots + 2\arctan\frac{2\lambda}{j+k-2} + \arctan\frac{2\lambda}{j+k}\,,
\end{eqnarray}
where  $M_k$ is the number of  $k$-strings present in a state such that $\sum_{k}kM_k=M$. The Takahashi quantum numbers, $I^j_\alpha$, which are  strictly non-repetitive,   are then  given by
\begin{eqnarray}\label{takahashi} 
\mid I_{\alpha}^j\mid \leq \frac{1}{2} \left(N-1-\sum_{k=1}\left[2 \mbox{min}(j,k)-\delta_{j,k}\right]M_k\right)\,.
\end{eqnarray}

%------------------------------------------------------------------------------------------------------------------------------------------------------------------------------------------
\section{Regularization for the singular solutions}
In this section we review  the regularization  of the singular solutions, which  was   introduced  in   \cite{vlad1}  and later pursued  in  detail in   \cite{bei,nepot,nepoh, nepo1,nepo,kirillov2}, as these results are essential in our study.  As mentioned in the introduction,  the singular sets of rapidities  make  the eigenvalues and the eigenvectors ill-defined.  It is manifest from the expression that  the Bethe  eigenstate  (\ref{vec1})  vanishes and the eigenvalue equation   (\ref{eigen})  diverges.   
By considering  typical  singular  solutions for the  $M$ down-spins as,
\begin{eqnarray}\label{sgn}
\Big\{\lambda_1=\frac{i}{2},\lambda_2= -\frac{i}{2}, \lambda_3,\lambda_4, \cdots, \lambda_M \Big\}\,,
\end{eqnarray} 
it can be  easily seen that   the presence of $\pm \frac{i}{2}$ in the singular solutions are responsible for  the pathology in  the expression of  the Bethe eigenstate  and the eigenvalue.
To handle this situation    the following  regularization   are used 
\begin{eqnarray}\label{regu}\nonumber
{\tilde\lambda}_1 &=& a\epsilon + \frac{i}{2}\left(1 + 2\epsilon^N\right)\,,\\
\tilde{\lambda}_2 &=& a\epsilon - \frac{i}{2}\left(1 + 2\epsilon^N\right)\,,
\end{eqnarray}
where $a$ is  a complex  constant  and $\epsilon$ is  a complex parameter, whose $\epsilon \to 0$ limit  gives the singular  solutions.  A rescaling of   $\epsilon$   by  $a$  reduces   eq.  (\ref{regu})  to the  one considered and extensively  discussed  in  \cite{nepo}.   In this respect see also eq. (31)  of ref. \cite{vlad1}  and   eq.  (3.4)  of  ref.  \cite{bei}, where the same  regularization   has been considered.

To obtain  the  conditions  for  
$\{ \lambda_3,\lambda_4, \cdots, \lambda_M \}$,    a  well-defined   Bethe state  with   $M$ rapidities  $\{ {\tilde\lambda}_\alpha\}$=$\{ {\tilde\lambda}_1, {\tilde\lambda}_2,  \lambda_3, \lambda_4, \cdots, \lambda_M \}$   of the form 
\begin{eqnarray}\label{wellvec}
|{\tilde\lambda}_1, {\tilde\lambda}_2, \lambda_3, \lambda_4, \cdots,\lambda_M\rangle= \frac{1}{(\tilde{\lambda}_1-\frac{1}{2}i)^N}\prod_{\alpha=1}^MB({\tilde\lambda}_\alpha)|\Omega\rangle \,,
\end{eqnarray}
is necessary. Action of $t(\lambda)$  on  (\ref{wellvec}) in   $\epsilon \to 0$ limit  is given by
\begin{eqnarray}\label{welltb}\nonumber
\lim_{\epsilon\to 0}t(\lambda)\frac{1}{(\tilde{\lambda}_1-\frac{1}{2}i)^N}\prod_{\alpha=1}^MB({\tilde\lambda}_\alpha)|\Omega\rangle &=& 
\lim_{\epsilon\to 0}\Lambda\left(\lambda, \{{\tilde\lambda}_\alpha\}\right) \frac{1}{(\tilde{\lambda}_1-\frac{1}{2}i)^N}\prod_{\alpha=1}^MB({\tilde\lambda}_\alpha)|\Omega\rangle + \\
&& \lim_{\epsilon\to 0}\sum_{k=1}^{M} 
\Lambda_{k}\left(\lambda, \{{\tilde\lambda}_\alpha\}\right)\frac{(\tilde{\lambda}_k-\frac{1}{2}i)^N}{(\tilde{\lambda}_1-\frac{1}{2}i)^N} B(\lambda)\prod_{\substack{{\alpha =1}\\{\alpha \neq k}}}^MB({\tilde\lambda}_\alpha)|\Omega\rangle\,,
\end{eqnarray}
where  
\begin{eqnarray}\label{wellteigen}
\lim_{\epsilon\to 0}\Lambda\left(\lambda, \{{\tilde\lambda}_\alpha\}\right)= \left(\lambda +\frac{i}{2}\right)^{N-1} \left(\lambda -\frac{3}{2}i\right)\prod_{\alpha=3}^M\frac{\lambda-\lambda_\alpha-i}{\lambda-\lambda_\alpha} + 
\left(\lambda - \frac{i}{2}\right)^{N-1} \left(\lambda + \frac{3}{2}i\right)\prod_{\alpha=3}^M\frac{\lambda-\lambda_\alpha+i}{\lambda-\lambda_\alpha}\,,
\end{eqnarray}
is   the eigenvalue of the transfer matrix  for the singular solutions  and   the  unwanted terms are 
\begin{eqnarray}\label{welltunw}
\lim_{\epsilon\to 0}\Lambda_k\left(\lambda, \{\tilde{\lambda}_\alpha\}\right)= 
\lim_{\epsilon\to 0} \frac{i}{\lambda-\tilde{\lambda}_k}\left[\frac{\left(\tilde{\lambda}_k + \frac{i}{2}\right)^N}{\left(\tilde{\lambda}_k - \frac{i}{2}\right)^N }\prod_{\substack{{\alpha =1}\\{\alpha\neq k}}}^M\frac{\tilde{\lambda}_k-\tilde{\lambda}_\alpha-i}{\tilde{\lambda}_k-\tilde{\lambda}_\alpha} -\prod_{\substack{{\alpha =1}\\{\alpha\neq k}}}^M\frac{\tilde{\lambda}_k-\tilde{\lambda}_\alpha+i}{\tilde{\lambda}_k-\tilde{\lambda}_\alpha}\right]\,,~ k= 1, 2, \cdots M\,.
\end{eqnarray}
Here we remark that both sides of (\ref{welltb})  have   finite and well-defined  limit and most importantly the  Bethe eigenstate  is  finite in the limit   and now  not a null state.
Note that  (\ref{welltb}) becomes the eigenvalue equation corresponding to  the singular solutions  if  the unwanted  terms  (\ref{welltunw})  vanish, i.e.
in the $\epsilon \to 0$ limit  the following equations   can be obtained 
\begin{eqnarray}\label{abethe1}
a^N &=&i^{N}\prod_{\alpha=3}^M\frac{\lambda_\alpha+\frac{1}{2}i}{\lambda_\alpha-\frac{3}{2}i}\,,\\
\label{abethe2}
a^N &=&(-i)^{N}\prod_{\alpha=3}^M\frac{\lambda_\alpha-\frac{1}{2}i}{\lambda_\alpha+\frac{3}{2}i}\,,\\
\label{abethe3}
\left(\frac{\lambda_\alpha- \frac{1}{2}i}{\lambda_\alpha+ \frac{1}{2}i}\right)^{N-1}&=& \frac{\lambda_\alpha- \frac{3}{2}i}{\lambda_\alpha+\frac{3}{2}i}\prod_{\substack{ {\beta=3}\\{\beta\neq \alpha}}}^{M}\frac{\lambda_\alpha-\lambda_\beta-i}{\lambda_\alpha-\lambda_\beta+i}\,,~~~  \alpha=3,4, \cdots, M\,.
\end{eqnarray}
Equating the two expressions  (\ref{abethe1})  and (\ref{abethe2}) one  obtains,   
\begin{eqnarray}\label{a1bethe1}
\prod_{\alpha=3}^M\frac{\lambda_\alpha+\frac{1}{2}i}{\lambda_\alpha-\frac{3}{2}i}&=&(-1)^{N}\prod_{\alpha=3}^M\frac{\lambda_\alpha- \frac{1}{2}i}{\lambda_\alpha+\frac{3}{2}i}\,.
\end{eqnarray}
Note that    eq. (\ref{abethe3})  was obtained in  \cite{sid,nepo}   and    eq.  (\ref{a1bethe1})  was obtained   in  \cite{nepo}.   One can  regard   the set of  eqs. (\ref{abethe3})-(\ref{a1bethe1})  as the Bethe ansatz equations for the singular solutions, as  the distinct and self-conjugate solutions  produce the well-defined   Bethe eigenstates and the eigenvalues for the singular solutions of  the transfer matrix and for the Hamiltonian $H$.   They   are in agreement with the  statement in   \cite{vlad2} that the distinct and self-conjugate  solutions of the Bethe equations  are    physical solutions.  For our purpose  we consider   eqs. (\ref{abethe3})-(\ref{a1bethe1})  to study  the  singular solutions for the even(odd) length chains. 

Taking the product of all the Bethe equations in  (\ref{abethe3})  one obtains 
\begin{eqnarray}\label{a1bethe1gn}
\prod^M_{\alpha = 3}  \left(\frac{\lambda_\alpha- \frac{1}{2}i}{\lambda_\alpha+ \frac{1}{2}i}\right)^{N-1} = \prod^M_{\alpha = 
3}\frac{\lambda_\alpha- \frac{3}{2}i}{\lambda_\alpha+\frac{3}{2}i}\,.
\end{eqnarray}
Dividing both sides of  eq. (\ref{a1bethe1gn})  by the both sides  of  eq   (\ref{a1bethe1})  the  condition  \cite{nepo}
\begin{eqnarray}\label{eqnw}
\left(- \prod_{i=3}^M \frac{\lambda_i + \frac{i}{2}}{\lambda_i -\frac{i}{2}}\right)^N=1\,,
\end{eqnarray}
can be obtained, as   also  pointed  out by Nepomechie   in a private commutation.
The set of      eqs. (\ref{abethe3})  and  (\ref{eqnw})   have been   considered   in \cite{nepo1}  to obtain   the physical singular solutions.  Here we   remark  that   in  ref.  \cite{bei} it  is has been addressed    that  the singular solutions  of even-length spin chains in  odd down-spin  sectors  satisfy     a trace  condition  (see eq. (2.4) and the related discussion  after eq. (3.4) of ref. \cite{bei}) in the    $\epsilon \to 0$ limit
\begin{eqnarray}\label{trace}
\lim_{\epsilon \to 0} \prod_{i=1}^M \frac{\tilde{\lambda}_i + \frac{i}{2}}{\tilde{\lambda}_i -\frac{i}{2}} =- \prod_{i=3}^M \frac{\lambda_i + \frac{i}{2}}{\lambda_i -\frac{i}{2}}=1\,,
\end{eqnarray}
The authors   assumed that the singular solutions  satisfying the trance condition  (\ref{trace})  are invariant under the sign changes of their rapidities.  In the odd-down spin sectors  the singular solutions then can be written in the form  $\{ \lambda_1=\frac{i}{2}, \lambda_2= -\frac{i}{2}, \lambda_3=0, \lambda_4, -\lambda_4, \cdots, \lambda_{M-1}, -\lambda_{M-1}\}$. Now note that  
$\{\lambda_3=0, \lambda_4, -\lambda_4, \cdots, \lambda_{M-1}, -\lambda_{M-1}\}$  automatically  satisfy   the trace condition  (\ref{trace}). 

As evident from eq (\ref{abethe1})-(\ref{abethe2}), the parameter $a$ is a function of the rapidities  $\{\lambda_\alpha, \alpha=3,4, \cdots, M\}$,  in general.    To  obtain the singular  Bethe eigenstates   for   $M$ down-spins   we  need  to  use 
\begin{eqnarray}\label{regu1}\nonumber
{\tilde\lambda}_1 &=& i\sqrt[N]{\prod_{\alpha=3}^M\frac{\lambda_\alpha+\frac{1}{2}i}{\lambda_\alpha-\frac{3}{2}i}}\epsilon + \frac{i}{2}\left(1 + 2\epsilon^N\right)\,,\\
\tilde{\lambda}_2 &=&  i\sqrt[N]{\prod_{\alpha=3}^M\frac{\lambda_\alpha+\frac{1}{2}i}{\lambda_\alpha-\frac{3}{2}i}}\epsilon - \frac{i}{2}\left(1 + 2\epsilon^N\right)\,,
\end{eqnarray}
in the  Bethe state  (\ref{wellvec})  and take the   $\epsilon\to 0$  limit
\begin{eqnarray}\label{vec}
|\lambda_1, \lambda_2,\cdots,\lambda_M\rangle= \lim_{\epsilon\to 0}\frac{1}{(\tilde{\lambda}_1-\frac{1}{2}i)^N}B(\tilde{\lambda}_1)B(\tilde{\lambda}_2)\prod_{\alpha=3}^MB(\lambda_\alpha)|\Omega\rangle \,.
\end{eqnarray}
The   eigenvalue of  the Hamiltonian $H$   for the  singular solution  $\{ \frac{i}{2}, - \frac{i}{2}, \lambda_3, \lambda_4, \cdots, \lambda_M\}$   can be obtained from the  eigenvalue eq. (\ref{wellteigen})  of the transfer matrix for the singular solutions   as  \cite{kirillov2}
\begin{eqnarray}\label{eigenr}
E=  \frac{J}{2}\left(-i\left[\frac{d}{d\lambda}\log \lim_{\epsilon\to 0}\Lambda\left(\lambda, \{{\tilde\lambda}_\alpha\}\right)\right]_{\lambda= -\frac{i}{2}}-N\right)=-J\left[ 1+\frac{1}{2}\sum_{\alpha=3}^{M}\frac{1}{\left(\lambda_\alpha^2+ \frac{1}{4}\right)}\right]\,.
\end{eqnarray}

%------------------------------------------------------------------------------------------------------------------------------------------------------------------------------------------
\section{Even Length Spin Chain} \label{sec4}
Numerically  the even length  spin-$1/2$ chain  has been investigated  for some finite values   of the length $N$ \cite{hag,nepo1}.  It has been observed numerically  that for the singular solutions  the rapidities are distributed symmetrically \cite{hag}.   Alternatively,  in the language of rigged configurations the singular solutions  of an  even-length spin chain  are flip invariant \cite{kirillov}.  Based on these, we  in our previous work  \cite{giri} assumed that the sum of the rapidities for the singular solutions of an even length spin-$1/2$ chain  vanishes.  Here, we discuss  this assumption  in the light of regularization  as well as the singular solutions  in general.  To start with,  let us consider the lowest down-spin sector for a singular solution to exist, i.e, $M=2$.  In this case   eq.  (\ref{regu1}) reduces to
\begin{eqnarray}\label{eregunepo}\nonumber
{\tilde\lambda}_1 &=& i\epsilon  + \frac{i}{2}\left(1 + 2\epsilon^N\right)\,,\\
\tilde{\lambda}_2 &=& i\epsilon  - \frac{i}{2}\left(1 + 2\epsilon^N\right)\,.
\end{eqnarray}
The Bethe eigenstate,  in this case,    takes a  simple form \cite{essler,nepo1} (see  {\bf Appendix  \ref{app1}} for the derivation)
\begin{eqnarray}\label{sgnvec1}
| \frac{i}{2},- \frac{i}{2}\rangle  \equiv  \sum_{j=1}^N (-1)^j S^-_jS^-_{j+1}|\Omega\rangle\,,
\end{eqnarray}
with the eigenvalue   $E=-J$.   We numerically confirmed  up to some lengths  of the spin chain that  eq. (\ref{sgnvec1})  is indeed the highest weight singular state.  For  $N=6$, it takes the form 
\begin{eqnarray}\label{sgnvec1n}
| \frac{i}{2},- \frac{i}{2}\rangle  \equiv  \left( 0_3, -1, 0_2, 1, 0_5, -1, 0_{11}, 1, 0_8, 1,  0_{14}, -1,  0_{15}  \right)\,,
\end{eqnarray}
where $0_m$ is the  short form of  $m$ consecutive $0$'s, for example $0_3= 0, 0, 0$. For the three down spin  sector, $M=3$,  eq (\ref{a1bethe1})  reads as 
\begin{eqnarray}\label{econd1}
\frac{\lambda_3+\frac{1}{2}i}{\lambda_3-\frac{3}{2}i}-\frac{\lambda_3-\frac{1}{2}i}{\lambda_3+\frac{3}{2}i}=0\,,
\end{eqnarray}
whose only solution  is $\lambda_3=0$ and it is also a solution  of  eq (\ref{abethe3}),  which means   $\{\lambda_1= \frac{i}{2},\lambda_2=- \frac{i}{2},\lambda_3=0\}$  is the only  solution  of the Bethe ansatz equations for the singular solutions  (\ref{abethe3})- (\ref{a1bethe1}).  
The regularization in this case becomes 
\begin{eqnarray}\label{regunepo3}\nonumber
{\tilde\lambda}_1 &=& i{\sqrt[N]{-\frac{1}{3}}}\epsilon + \frac{i}{2}\left(1 + 2\epsilon^N\right)\,,\\
\tilde{\lambda}_2 &=&i{\sqrt[N]{-\frac{1}{3}}}\epsilon  - \frac{i}{2}\left(1 + 2\epsilon^N\right)\,.
\end{eqnarray}
The Bethe eigenstate,  in this case,    becomes (see  {\bf Appendix \ref{app2}} for the  derivation)
\begin{eqnarray}\label{sgnvec12}
| \frac{i}{2},- \frac{i}{2},0\rangle  \equiv  \sum_{j=1}^N (-1)^j S^-_jS^-_{j+1} \left(\sum_{k=1}^N (-1)^k S^-_k\right)|\Omega\rangle\,,
\end{eqnarray}
with the eigenvalue   $E=-3J$.   We numerically confirmed up to some lengths  that  eq. (\ref{sgnvec12})  is indeed the highest weight singular Bethe eigenstate.  For  $N=6$, it takes the form 
\begin{eqnarray}\label{sgnvec12n}
| \frac{i}{2},- \frac{i}{2}, 0 \rangle  \equiv  \left( 0_{11}, 1, 0, -1, 0_5, -1, 0_{2}, 1, 0_2, 1, -1,   0_{10}, 1, -1,  0_{2}, -1, 0_2,  1, 0_5, 1, 0, -1, 0_{11}  \right)\,.
\end{eqnarray}
Analytic calculation for $M \geq 4$ becomes  more  difficult, but we can still proceed to find a   symmetry, which   the rapidities  for the singular solutions follow. 
For general values of $M$    eq.  (\ref{a1bethe1})  reads as
\begin{eqnarray}\label{econd2}
\prod_{\alpha=3}^M\frac{\lambda_\alpha+\frac{1}{2}i}{\lambda_\alpha-\frac{3}{2}i}-\prod_{\alpha=3}^M\frac{\lambda_\alpha-\frac{1}{2}i}{\lambda_\alpha+\frac{3}{2}i}=0\,.
\end{eqnarray}
If a set of rapidities satisfy the conditions 
\begin{eqnarray}\label{econd3}
\lambda_\alpha + \lambda_{\alpha+1}=0\,, ~~~~~~~ \mbox{for}~~  \alpha= 3, 5,  \cdots,  M-1\,,
\end{eqnarray}
for even $M$,  then they satisfy   eq. (\ref{econd2}).  Similarly,  if a set of rapidities  satisfy the conditions 
\begin{eqnarray}\label{econd4}
\nonumber
\lambda_3 &=& 0\,,\\
\lambda_\alpha + \lambda_{\alpha+1} &=& 0\,, ~~~\mbox{for}~~  \alpha= 4, 6,  \cdots,  M-1\,.
\end{eqnarray}
for odd $M$, then they  satisfy  eq. (\ref{econd2}).
It follows   that the singular solutions $\{\lambda_\alpha\}= \{\lambda_1, \lambda_2, \cdots, \lambda_M\}$, which  satisfy  the   conditions  (\ref{econd3}) or (\ref{econd4}), are    are invariant  under the sign changes of each of  their  rapidities  i.e.   $\{\lambda_\alpha\}= \{-\lambda_\alpha\}$. It   implies   that   the sum of   rapidities  of such a   singular solution for even $N$  vanishes \cite{giri}, i.e,
\begin{eqnarray}\label{econd5}
\sum_{\alpha=1}^M\lambda_\alpha=0\,.
\end{eqnarray}
In the language of rigged  configurations  the  conditions   (\ref{econd3}) and  (\ref{econd4})  or   the condition  (\ref{econd5})  is equivalent to the flip  invariance of the riggings,  which has  to  be satisfied by singular solutions according to a conjecture in \cite{kirillov}.
Note that  the conditions  (\ref{econd3}) reduce  the  Bethe ansatz equations for the singular solutions  in  an even down-spin  sector   to a system of equations  of   $\left(M-2\right)/2$  rapidities   $\{\lambda_3, \lambda_5, \cdots, \lambda_{M-1}\}$
\begin{eqnarray}\label{e2bethe1}
\left(\frac{\lambda_\alpha- \frac{1}{2}i}{\lambda_\alpha+ \frac{1}{2}i}\right)^{N-2} &=& \frac{\lambda_\alpha- \frac{3}{2}i}{\lambda_\alpha+\frac{3}{2}i} \prod_{\beta\neq \alpha}^{M-1}\frac{\lambda_\alpha-\lambda_\beta-i}{\lambda_\alpha-\lambda_\beta+i}
\frac{\lambda_\alpha+\lambda_\beta-i}{\lambda_\alpha+\lambda_\beta+i}\,, ~~~  \alpha, \beta \in [3, 5, \cdots, M-1]\,.
\end{eqnarray}
Similarly  the conditions   (\ref{econd4}) reduce  the  Bethe ansatz equations for the singular solutions  in an odd down-spin sector   to a system of equations  of   $\left(M-3\right)/2$  rapidities   $\{\lambda_4, \lambda_6, \cdots, \lambda_{M-1}\}$
\begin{eqnarray}\label{e2bethe2}
\left(\frac{\lambda_\alpha- \frac{1}{2}i}{\lambda_\alpha+ \frac{1}{2}i}\right)^{N-2} &=& \frac{\lambda_\alpha- \frac{3}{2}i}{\lambda_\alpha+\frac{3}{2}i} \frac{\lambda_\alpha- i}{\lambda_\alpha+i} \prod_{\beta\neq \alpha}^{M-1}\frac{\lambda_\alpha-\lambda_\beta-i}{\lambda_\alpha-\lambda_\beta+i}
\frac{\lambda_\alpha+\lambda_\beta-i}{\lambda_\alpha+\lambda_\beta+i}\,, ~~~  \alpha, \beta \in [4, 6, \cdots, M-1]\,.
\end{eqnarray}

One can numerically show that  apart from solutions of the form  (\ref{econd5}) there are no other  solutions  for even length chains up to,  for instance $N=14$.

%-------------------------------------------------------------------------------------------------------------------------------------------------------------------------------------

\section{Odd Length Spin Chain} \label{sec5}
Singular solutions for  the odd length chain is not much discussed in the literature until  very recently   \cite{nepo1,nepo}.   
For two down-spins,  $M=2$,  left hand side of eq. (\ref{a1bethe1}) is given by  $+ 1$, while the right hand side is given by  $- 1$.  The disagreement between both sides  implies   that there is no singular solution.   
For three down-spins,  $M=3$,  we obtain  from  eq.  (\ref{a1bethe1})
\begin{eqnarray}\label{ocond1}
\frac{\lambda_3+\frac{1}{2}i}{\lambda_3-\frac{3}{2}i}+\frac{\lambda_3-\frac{1}{2}i}{\lambda_3+\frac{3}{2}i}=0\,,
\end{eqnarray}
whose solutions are  $\lambda_3= \pm \frac{\sqrt{3}}{2}$.  In order for them  to become  the Bethe roots  they  also   have to  satisfy  (\ref{abethe3}),  which  in this case  becomes 
\begin{eqnarray}\label{bethe2}
\left(\frac{\lambda_3- \frac{1}{2}i}{\lambda_3+ \frac{1}{2}i}\right)^{N-1}= \frac{\lambda_3- \frac{3}{2}i}{\lambda_3+\frac{3}{2}i} \,.
\end{eqnarray}
In   {\bf Appendix \ref{app3}} we derive from  (\ref{bethe2}) the following equation   for the singular solutions
\begin{eqnarray}\label{oddfac1}
\left(\lambda_3^2- \frac{3}{4}\right) \left[ \left(\lambda_3- \frac{i}{2}\right)^N - \left(\lambda_3 + \frac{i}{2}\right)^N + 4 i \lambda_3^2\sum_{r=0}^{\frac{N-3}{2}}
\lambda_3^{N-3-2r}\left(\frac{3}{4}\right)^r\left(\sum_{s=0}^{r}\left(-\frac{1}{3}\right)^s \Comb{N}{2s}\right)\right]= 0\,,
\end{eqnarray}
for  $N$  satisfying    
\begin{eqnarray}\label{oddfac2}
N=3\left(2k+1\right),  ~~~k =1, 2, 3,  \cdots\,.
\end{eqnarray}
We see  that  $\lambda_3= \pm \frac{\sqrt{3}}{2}$  are  indeed   solutions of  eq (\ref{oddfac1}), provided  the length $N$ of  the chain satisfies  eq. (\ref{oddfac2}).   Note that numerical evidence  for  (\ref{oddfac2})    has  already  been   found in \cite{nepo1}.
The regularization in this case becomes
\begin{eqnarray}\label{regunepo}\nonumber
{\tilde\lambda}_1 &=& i\sqrt[N]{\pm i\frac{1}{\sqrt{3}}}\epsilon + \frac{i}{2}\left(1 + 2\epsilon^N\right) \,,\\
\tilde{\lambda}_2 &=& i\sqrt[N]{\pm i \frac{1}{\sqrt{3}}}\epsilon  - \frac{i}{2}\left(1 + 2\epsilon^N\right)\,,
\end{eqnarray}
where $\pm$ correspond to the  regularization of the  two roots $\lambda_3= \pm \frac{\sqrt{3}}{2}$ respectively.
The Bethe eigenstates,  in this case,    become (see  {\bf Appendix \ref{app21}} for the derivation)
\begin{eqnarray}\label{sgnveco12}
| \frac{i}{2},- \frac{i}{2}, \pm \frac{\sqrt{3}}{2}\rangle \equiv \sum_{k=1}^N \left( \exp{(\pm \frac{\pi}{3}ki)} S^-_k \sum_{j=1}^N (-1)^{j+H(j-k)} S^-_jS^-_{j+1}\right)|\Omega\rangle\,,
\end{eqnarray}
with the eigenvalue   $E=-1.5J$.  For  $N=9$, numerically  we obtain the singular eigenstates  for  $\lambda_3= \pm \frac{\sqrt{3}}{2}$ as 
\begin{eqnarray} \nonumber
| \frac{i}{2},-\frac{i}{2},  \frac{\sqrt{3}}{2} \rangle  \equiv && \left( 0_{7}, 1, 0_3,  a, 0,  a,  a^*, 0_4,  -1,  0_2, 1,  0_2,  -a,  1, 0,  a,  0_6,  a^*, 0_2,  - a^*,  0_5,  a^*,  0_4,  a,  -1, 0,  a^*,  0_3, \right. \\ \nonumber
 &&\left. 1,  0_{10},  -a,  0_2,  a,  0_5, -a,  0_{11},  a,  0_8,  -a,  1, 0,  -a^*,  0_3, a,  0_7,  a^*,  0_{18},  1,  0_2,  -1,  0_5, 1,  0_{11},  -1,  0_{23},   \right. \\ \nonumber
&& \left.  1,  0_{16},  a,  -1, 0,  a^*,  0_3,  -a,  0_7, 1,  0_{15}, a,  0_{34},  a,  0,  a^*,  a^*,  0_2, -a,  0_2,  -a^*,  0_4,  1,  0_6,  a^*,  0_8,  \right. \\  \label{sgnveco12n}
&& \left.   -a^*,  0_{14},  -a^*,  0_{16},  a,  0_{30},  a^*,  0_{32},  a^*,  1, 0, -a^*,  0_3,  a,  0_7,  -1,  0_{15},  a^*,  0_{31},  1,  0_{63}  \right)\,, \\ \nonumber
| \frac{i}{2},-\frac{i}{2},  -\frac{\sqrt{3}}{2} \rangle  \equiv  &&  \left( 0_{7}, -a,   0_3,  -1, 0, -1, -a^*, 0_4,  a,  0_2, -a,  0_2, 1, -a, 0, -1, 0_6, -a^*, 0_2,  a^*, 0_2, 0_3, -a^*, 0_2, \right.\\ \nonumber
&& \left.  0_2, -1, a, 0, -a^*, 0_3, -a,  0_{10}, 1, 0_{2}, -1,  0_5, 1,  0_{11}, -1, 0_8,  1, -a,  0,   a^*, 0_3, -1,  0_7, -a^*,  0_{18}, \right. \\ \nonumber
&& \left.  -a,  0_2,  a,  0_5,  -a,  0_{11},  a,  0_{23},  -a,  0_{16}, -1,  a, 0,  -a^*,  0_3, 1,  0_7,  -a,  0_{15}, -1,  0_{34}, -1, 0, -a^*, \right. \\ \nonumber
&& \left.  -a^*,  0_2, 1,  0_2,  a^*,  0_4,  -a,  0_6,  -a^*, 0_8,  a^*,  0_{14},  a^*,  0_{16}, -1,  0_{30},  -a^*,  0_{32},  -a^*,  -a,  0,  a^* \right. \\ \label{sgnveco12n1}
 && \left.  0_3,  -1,  0_7,  a,  0_{15},  -a^*,  0_{31},  -a,  0_{63}\right)\,,
\end{eqnarray}

where  $a = -  \exp{( \frac{\pi}{3}i)}$.
%----------------------------------------------------------------------------------------------------
\begin{figure}[h!]\label{fig1}
  \centering
    \includegraphics[width=0.40\textwidth]{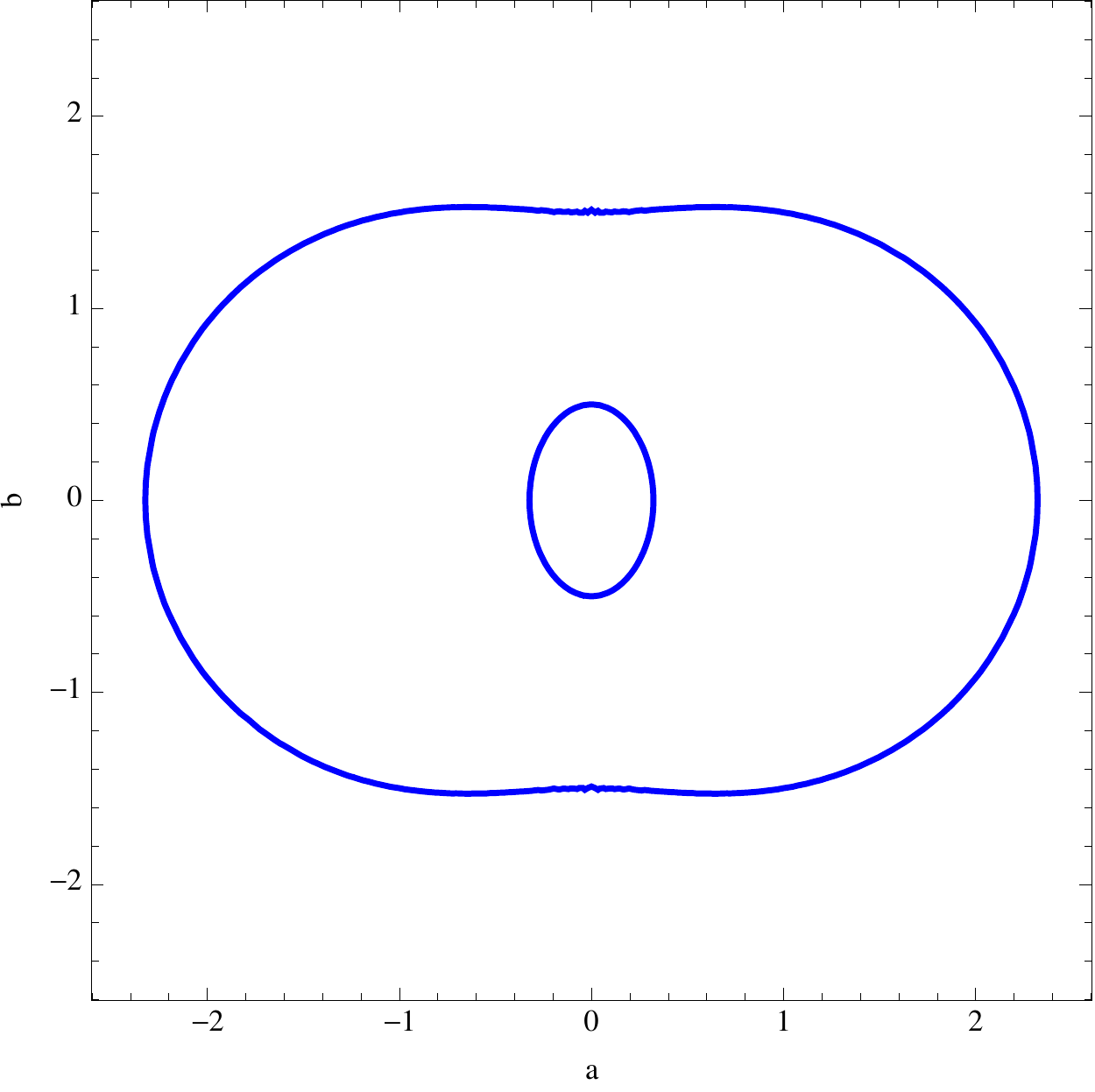}
  \caption{(Color online)Plot of  eq.  (\ref{ocond2}). We assign $a$ in the horizontal  axis and $b$ in the vertical axis.} \label{fg1}

\end{figure}
%----------------------------------------------------------------------------------------------------

%----------------------------------------------------------------------------------------------------
\begin{figure}[h!]
  \centering
    \includegraphics[width=0.40\textwidth]{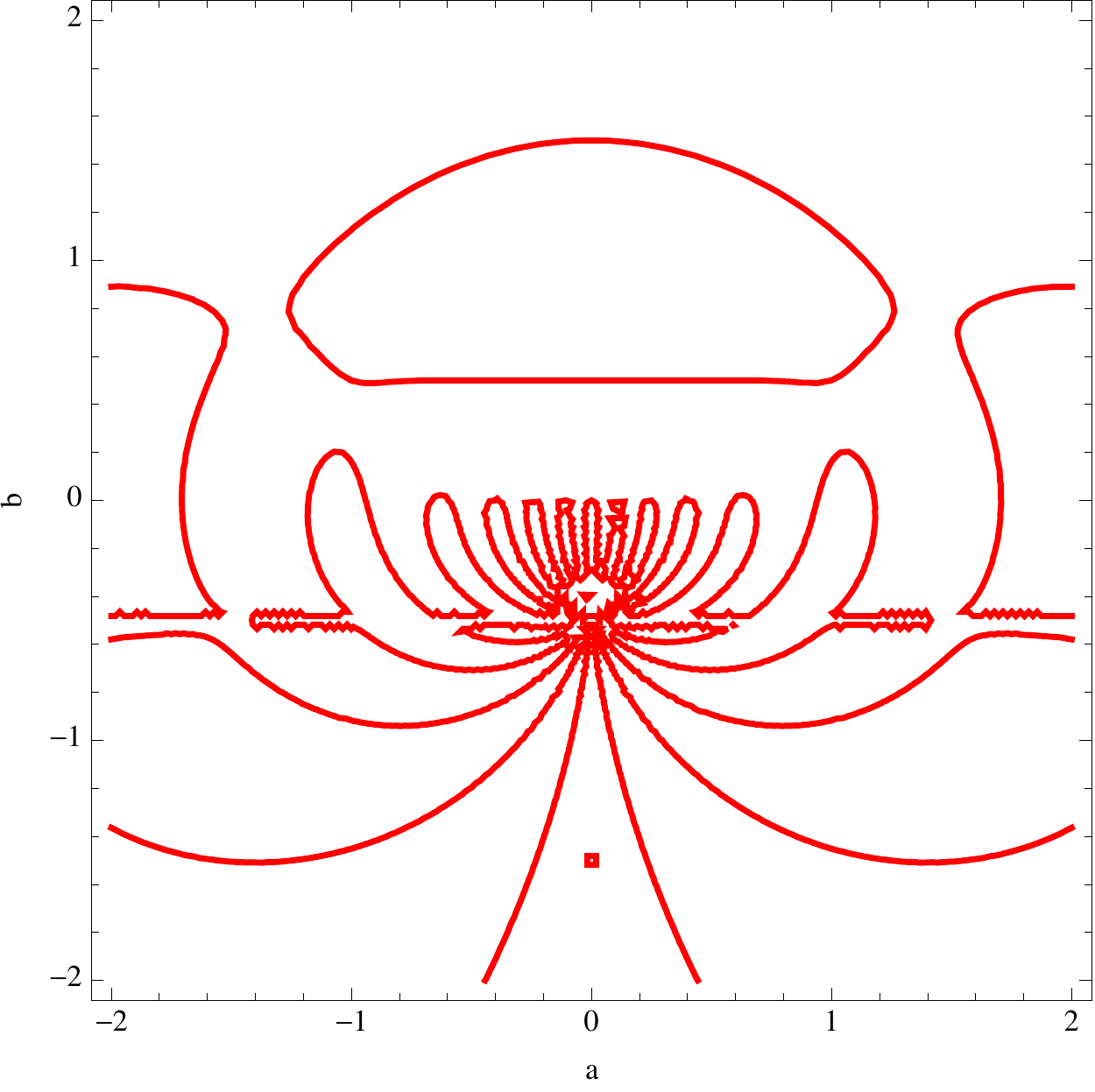}
  \caption{(Color online)Plot  eq.  (\ref{ocond3}) for $N=15$.} \label{fg2}

\end{figure}
%----------------------------------------------------------------------------------------------------

%----------------------------------------------------------------------------------------------------
\begin{figure}[h!]
  \centering
    \includegraphics[width=0.40\textwidth]{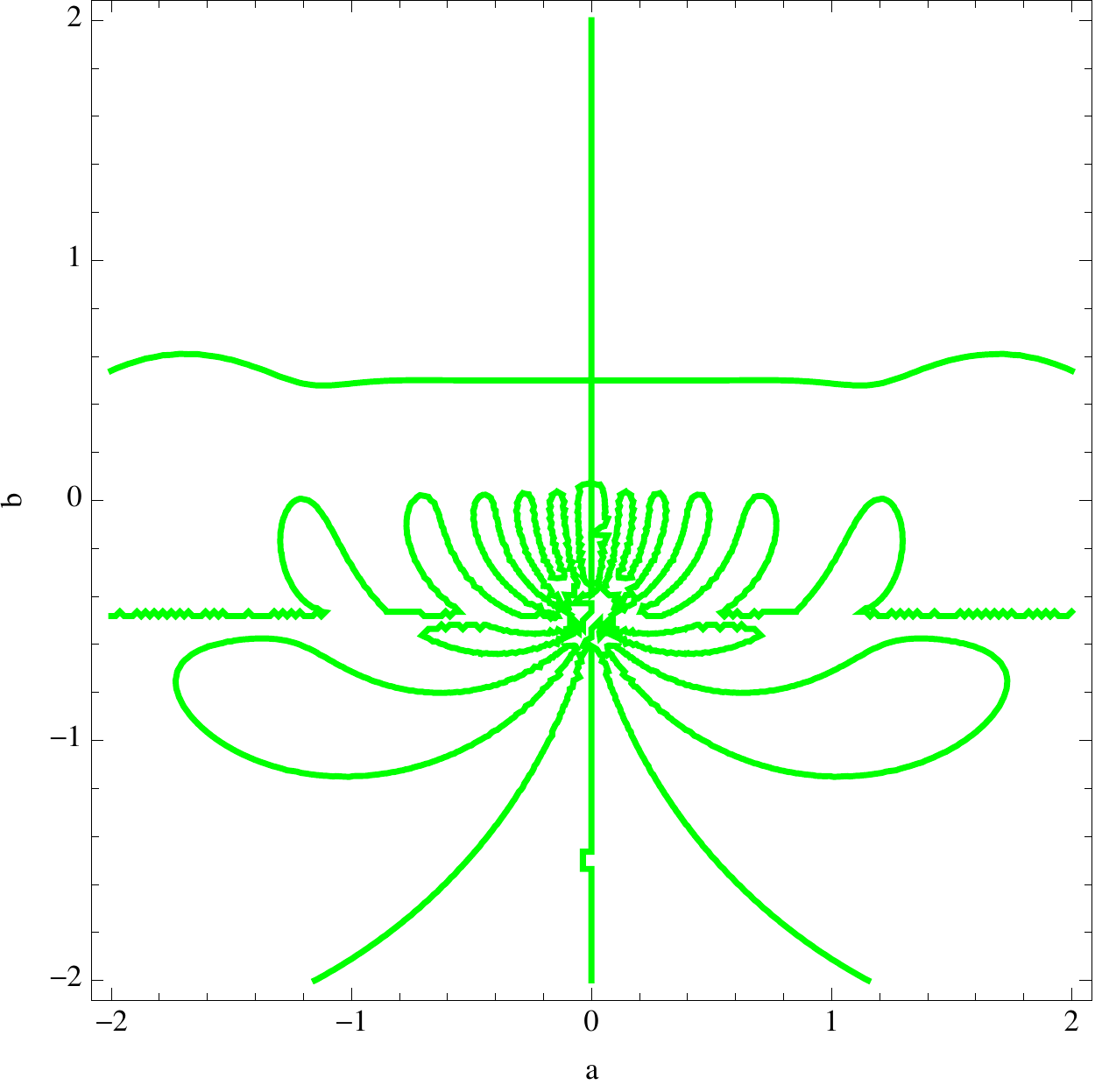}
  \caption{(Color online)Plot  eq.  (\ref{ocond4}) for $N=15$.}\label{fg3}

\end{figure}
%----------------------------------------------------------------------------------------------------

%----------------------------------------------------------------------------------------------------
\begin{figure}[h!]
  \centering
    \includegraphics[width=0.40\textwidth]{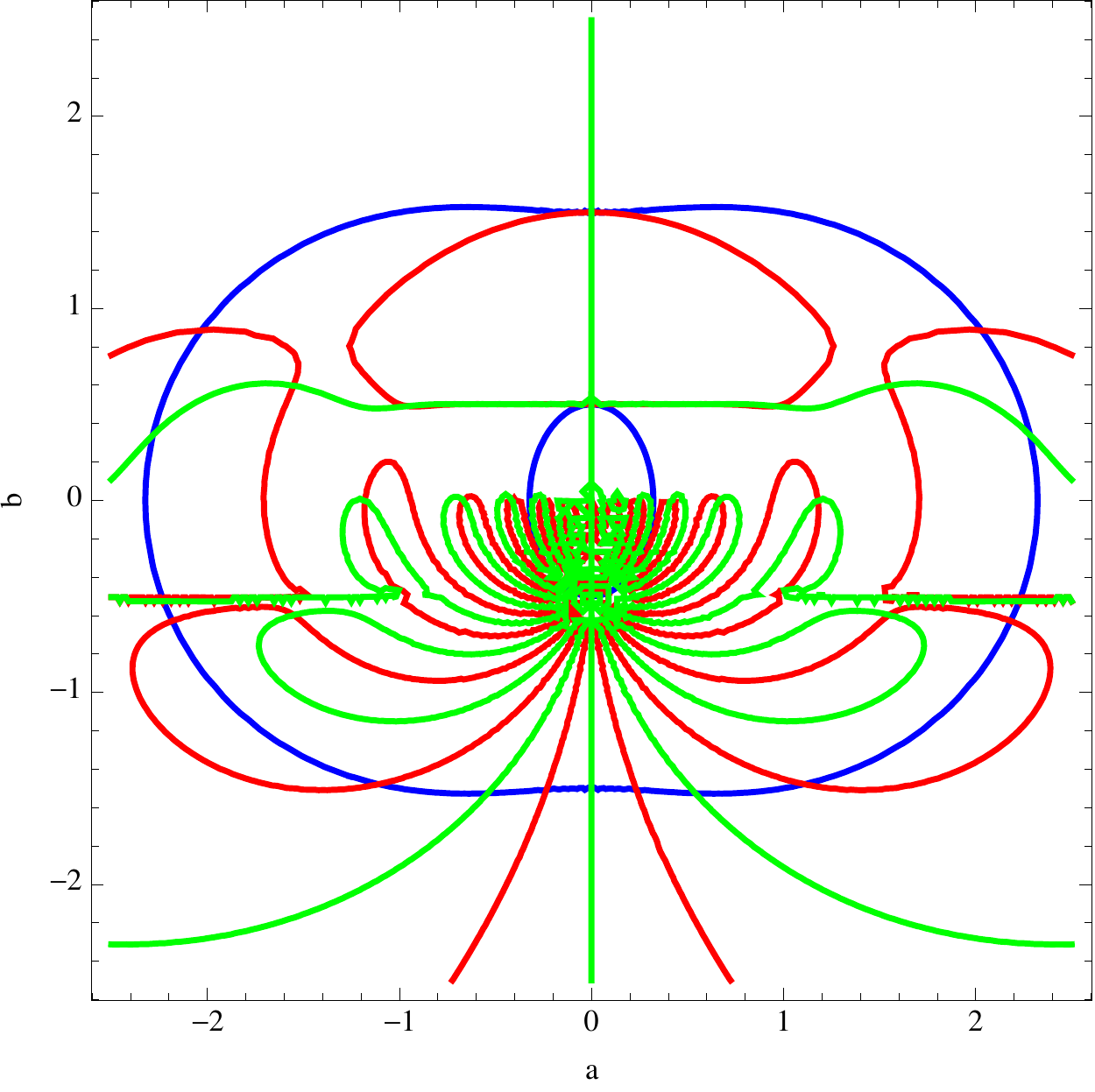}
  \caption{(Color online)Plots of  eqs.  (\ref{ocond2}) (blue curves),   (\ref{ocond3}) (red curves)  and   (\ref{ocond4}) (green curves)  for $N=15$.}\label{fg4} 

\end{figure}
%----------------------------------------------------------------------------------------------------

%----------------------------------------------------------------------------------------------------
\begin{figure}[h!]
  \centering
    \includegraphics[width=0.40\textwidth]{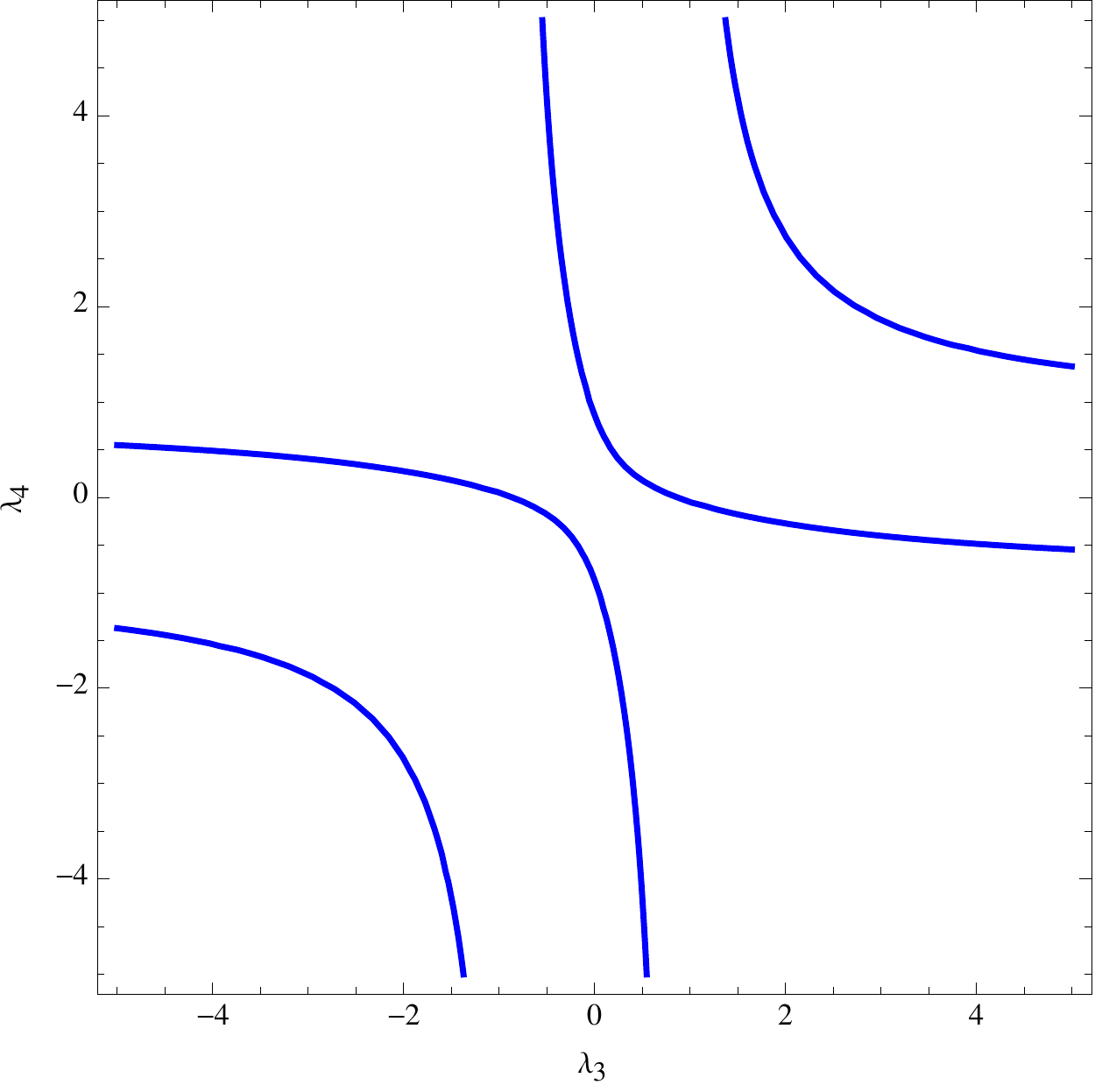}
  \caption{(Color online)Plot of  eq.  (\ref{ocond5}). We assign $\lambda_3$ in the horizontal axis and $\lambda_4$ in the vertical axis.}\label{fg5}

\end{figure}
%----------------------------------------------------------------------------------------------------

%----------------------------------------------------------------------------------------------------
\begin{figure}[h!]
  \centering
    \includegraphics[width=0.40\textwidth]{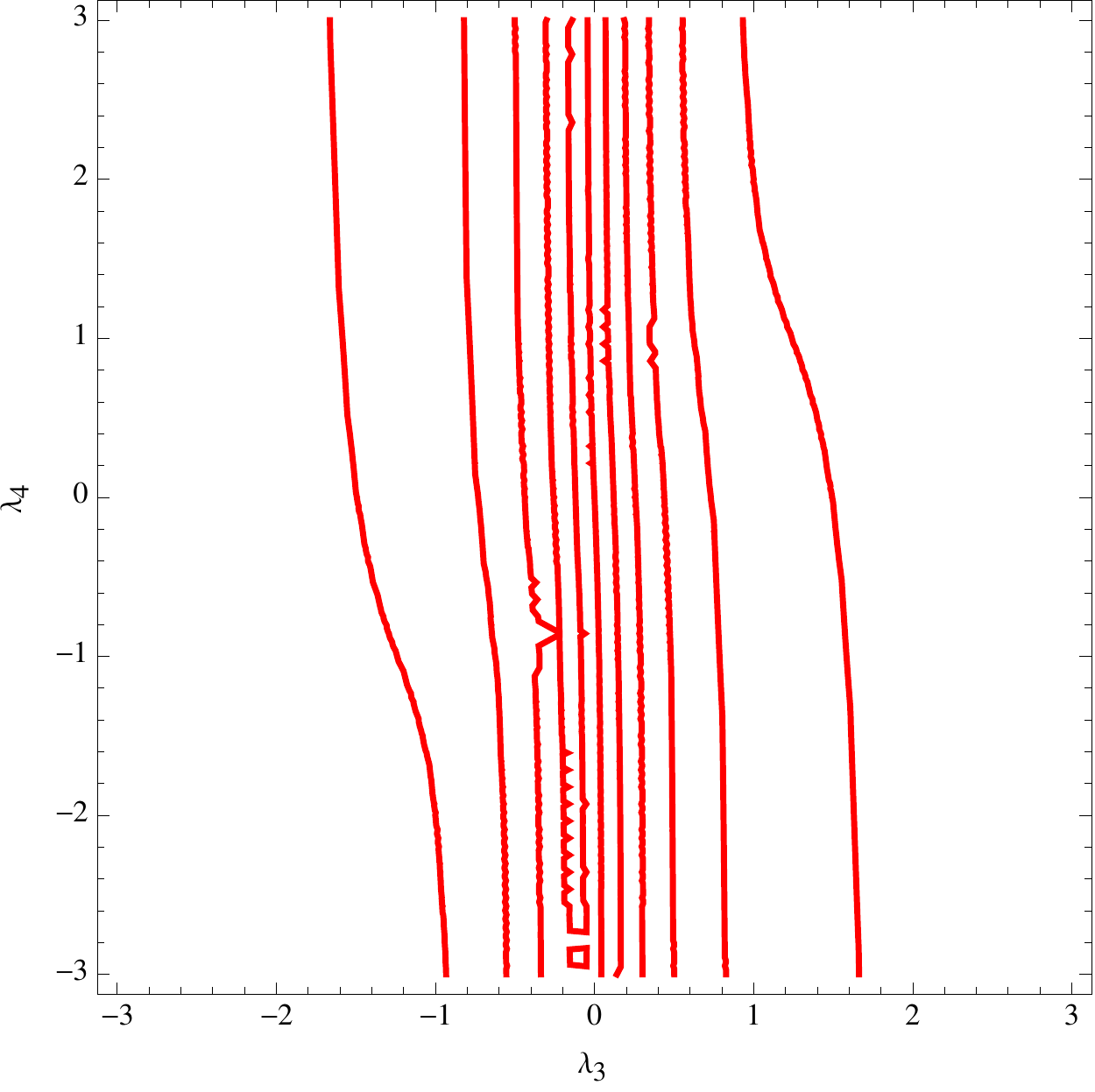}
  \caption{(Color online)Plot of  eq.  (\ref{ocond6}) for $N=15$.}\label{fg6}

\end{figure}
%----------------------------------------------------------------------------------------------------

%----------------------------------------------------------------------------------------------------
\begin{figure}[h!]
  \centering
    \includegraphics[width=0.40\textwidth]{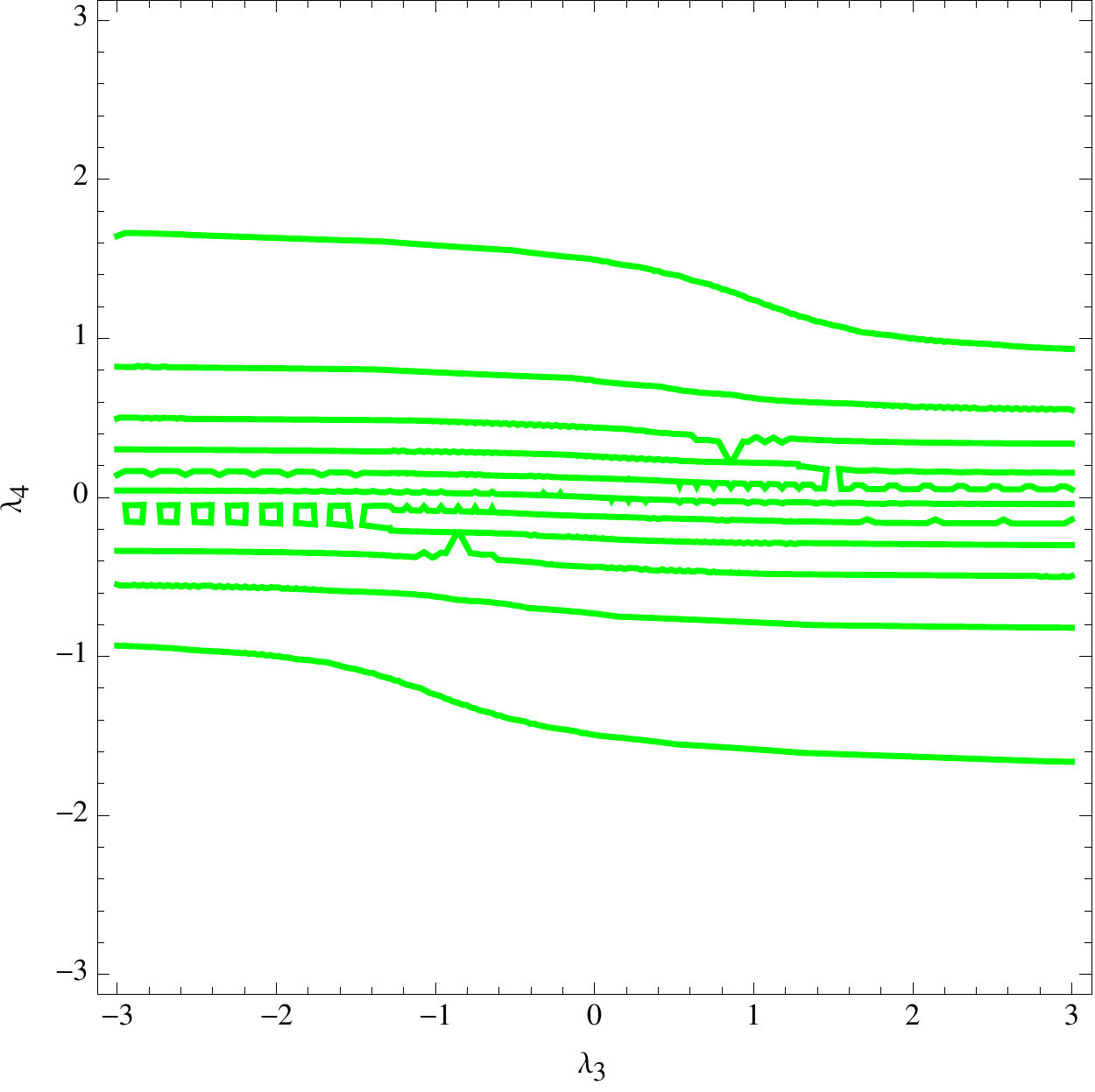}
  \caption{(Color online)Plot of  eq.  (\ref{ocond7}) for $N=15$.}\label{fg7}

\end{figure}
%----------------------------------------------------------------------------------------------------

%----------------------------------------------------------------------------------------------------
\begin{figure}[h!]
  \centering
    \includegraphics[width=0.40\textwidth]{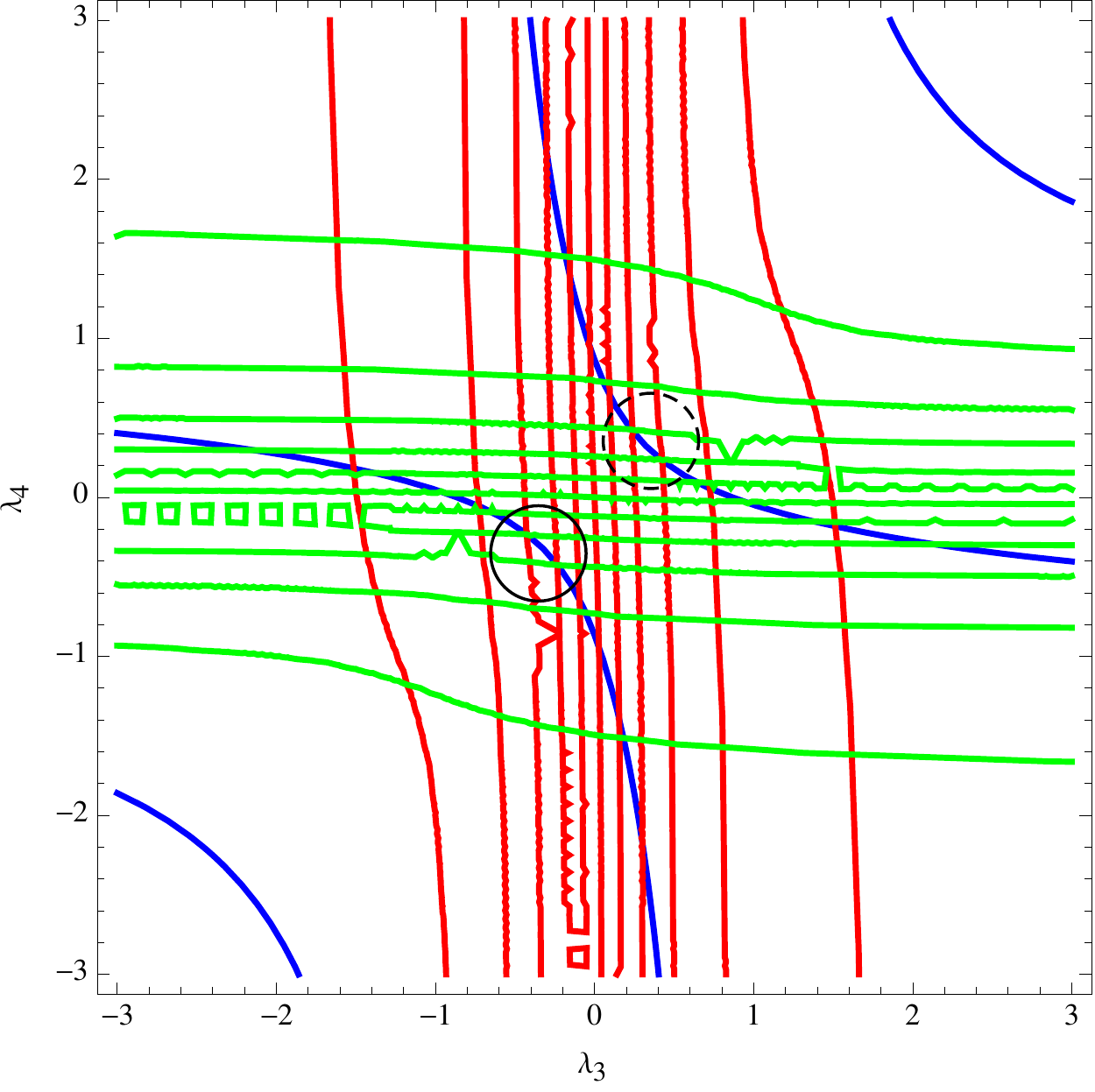}
  \caption{(Color online)Plot of  eqs.  (\ref{ocond5}) (blue curves), (\ref{ocond6}) (red curves) and (\ref{ocond7}) (green curves)   for $N=15$.}\label{fg8}

\end{figure}
%----------------------------------------------------------------------------------------------------

%----------------------------------------------------------------------------------------------------
\begin{figure}[h!]
  \centering
    \includegraphics[width=0.40\textwidth]{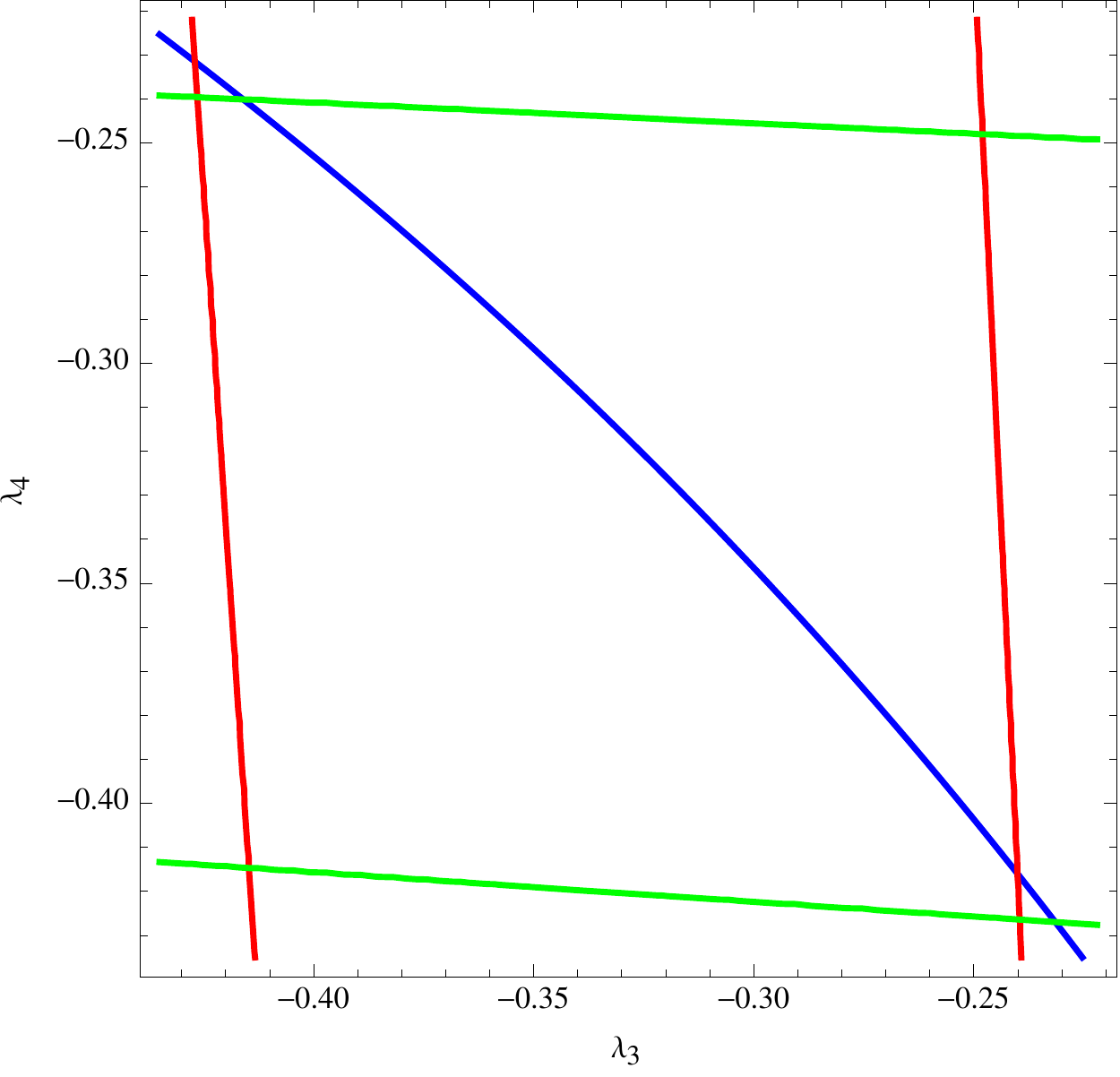}
  \caption{(Color online)Plot of the region inside solid circle in FIG.  \ref{fg8}.}\label{fg9}

\end{figure}
%----------------------------------------------------------------------------------------------------

%----------------------------------------------------------------------------------------------------
\begin{figure}[h!]
  \centering
    \includegraphics[width=0.40\textwidth]{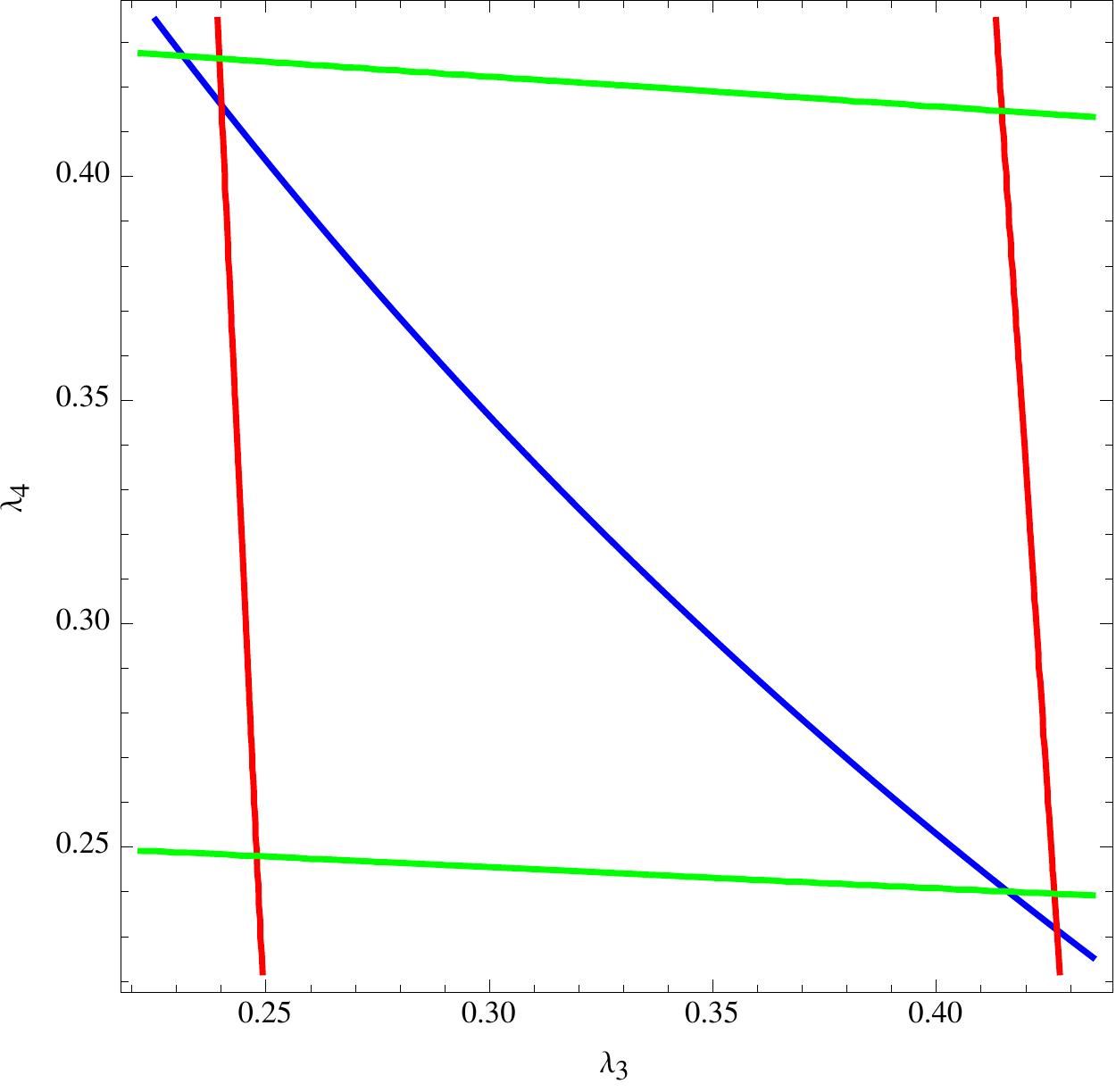}
  \caption{(Color online)Plot of the region inside  dashed circle in FIG. \ref{fg8}.}\label{fg10}

\end{figure}
%----------------------------------------------------------------------------------------------------

In the four down-spin sector, it is more difficult to  analytically  search for any possible   singular solutions.  One possible method  is to throughly look for all numerical  roots of the spin-$1/2$ XXX chain,  as  done   in \cite{nepo1}, who  obtained    no singular solutions for odd lengths   up to $N=13$.  However, a more efficient way would be just to concentrate on singular solutions. Here we  just plot  the graph associated with the Bethe ansatz equations for the singular solutions
and look for any possible  intersections of the  curves.   As an example we consider  the   $N=15$  case   but  it can  also  be  extended  to  other values of $N$.  There are two possible situations,  either   $\lambda_3$ and $\lambda_4$ are real  or they are complex conjugate to each other. 
Let us first discuss the complex  rapidity case.  Replacing  $\lambda_3=a+ib, \lambda_4=a-ib$ in   (\ref{a1bethe1}) we obtain 
\begin{eqnarray}\label{ocond2}
a^4+b^4+2a^2b^2-\frac{11}{2}a^2-\frac{5}{2}b^2+\frac{9}{16}=0\,,
\end{eqnarray}
%\begin{eqnarray}\label{ocond2}
%\frac{a +\left(b+\frac{1}{2}\right)i}{a +\left(b-\frac{3}{2}\right)i} \frac{a -\left(b-\frac{1}{2}\right)i}{a -\left(b+\frac{3}{2}\right)i}+
%\frac{a +\left(b-\frac{1}{2}\right)i}{a +\left(b+\frac{3}{2}\right)i} \frac{a -\left(b+\frac{1}{2}\right)i}{a -\left(b-\frac{3}{2}\right)i}=0
%\end{eqnarray}
which is plotted in FIG \ref{fg1}.  The other two equations  (\ref{abethe3})  are just the complex conjugate to each other, so we   equate the real part and the imaginary  part of both the sides 
\begin{eqnarray}\label{ocond3}
\mbox{Re}\left(\left[\frac{a + \left(b- \frac{1}{2}\right)i}{a+ \left(b+ \frac{1}{2}\right)i}\right]^{N-1}\right) = \mbox{Re}\left(\frac{a+ \left(b- \frac{3}{2}\right)i}{a +\left(b+\frac{3}{2}\right)i} \frac{2b-1}{2b+1}\right)\,,\\ \label{ocond4}
\mbox{Im}\left(\left[\frac{a + \left(b- \frac{1}{2}\right)i}{a+ \left(b+ \frac{1}{2}\right)i}\right]^{N-1}\right) = \mbox{Im}\left(\frac{a+ \left(b- \frac{3}{2}\right)i}{a +\left(b+\frac{3}{2}\right)i} \frac{2b-1}{2b+1}\right)\,.
\end{eqnarray}
Eq. (\ref{ocond3}) and   (\ref{ocond4}) are plotted for $N=15$ in  FIG. \ref{fg2} and FIG. \ref{fg3} respectively.  In order to have a solution, the    three curves   (\ref{ocond2})- (\ref{ocond4})  have to coincide at   complex conjugate points. From FIG \ref{fg4}, we see that the these  curves indeed  coincide    at   $(a=0, b=\pm \frac{1}{2})$, but  they  are  not  physical solutions, since the physical solutions for the spin-$1/2$ chain have to be  distinct.   For  $(a=0, b=\pm \frac{3}{2})$ although it seems from  FIG. \ref{fg4}  that  there are  intersections of the curves but they  actually do  not intersect.  Because,    although $(a=0, b=\pm \frac{3}{2})$ are  solutions of  eq. (\ref{ocond2}),    they are not  solutions of    (\ref{ocond3}) or  (\ref{ocond4}). It   can be easily seen that the right hand side of  both the equations either vanish or become infinity  while  the left hand side    is finite.   

Let us now consider the case when the two  rapidities  $\lambda_3, \lambda_4$ are real. From  eq. (\ref{a1bethe1}) we obtain
\begin{eqnarray}\label{ocond5}
\lambda_3^2\lambda_4^2-\frac{3}{4} \left(\lambda_3^2+\lambda_4^2\right) +\frac{9}{16}-4\lambda_3\lambda_4=0\,,
\end{eqnarray}
which does not have any real solutions of the form  $\lambda_3= -\lambda_4$, which is also  evident from FIG. \ref{fg5}.   The other two  equations obtained from   (\ref{abethe3})   are 
\begin{eqnarray}\label{ocond6}
\left(\lambda_3-\frac{1}{2}i\right)^{N-1} \left(\lambda_3 + \frac{3}{2}i\right)\left(\lambda_3 -\lambda_4+ i\right)-\left(\lambda_3+\frac{1}{2}i\right)^{N-1} \left(\lambda_3 - \frac{3}{2}i\right)\left(\lambda_3 -\lambda_4- i\right)=0\,,\\  \label{ocond7}
\left(\lambda_4-\frac{1}{2}i\right)^{N-1} \left(\lambda_4 + \frac{3}{2}i\right)\left(\lambda_4 -\lambda_3+ i\right)-\left(\lambda_4+\frac{1}{2}i\right)^{N-1} \left(\lambda_4 - \frac{3}{2}i\right)\left(\lambda_4 -\lambda_3- i\right)=0\,,
\end{eqnarray}
where  the real part of the first term  cancels with the real part of the second term in the left hand side of  both the above equations, while the imaginary part survives.  In  FIGs.  \ref{fg6}  and  \ref{fg7} we plot    eq. (\ref{ocond6}) and   (\ref{ocond7}),  respectively.  In FIG. \ref{fg8}   the  eqs.    (\ref{ocond5}),  (\ref{ocond6}) and   (\ref{ocond7})   have been plotted to see if there are any intersection of the three plots.    The  two regions  inside the solid and dashed circles  seem to have  intersection points.  However,  the region inside the solid circle    plotted in   FIG. \ref{fg9} and  the region inside the dashed circle plotted  FIG. \ref{fg10}  clearly show  that there is no intersection point at all.

%------------------------------------------------------------------------------------------------------------------------------------------------------------------------------------------

\section{Conclusions}
It is  known that the singular solutions  of the Bethe ansatz equations produce  ill-defined  Bethe eigenstates and  eigenvalues in the standard approach. Therefore, one needs to properly  regularize   the    solutions.  We in this paper are particularly interested in the  implications of this regularization on the  Bethe eigenstates for the even(odd) length spin chains  in some fixed down-spin sectors.   Specifically,  the analytic forms of the Bethe eigenstates for three down-spin sector of even and odd length spin  chains   have been obtained and  their    numerical  forms  in some fixed length chains are given. 
For the singular solutions  if  the   rapidities   are symmetrically distributed   in the complex plane  i.e.   $\{\lambda_\alpha\}= \{-\lambda_\alpha\}$  for an  even length spin-$1/2$ XXX chain then  the Bethe equations    are expressed  in a significantly  reduced form.   These equations can be handled  easily in the numerical process.     
We have  analytically  shown that in the  three down-spin  sector of the odd-length chain, there exist  singular  solutions for any  finite length of the spin chain of the form of   $N= 3\left(2k+1\right)$ with  $k = 1, 2, 3, \cdots$.   Searching for  any possible singular solutions for the   four down-spin sector  of  an odd-length chain is  more difficult.  However, we   have shown with an example of  $N=15$ that  it can be done easily  by simply plotting the  Bethe ansatz equations for the singular solutions and looking  for any possible intersections of the three curves.  For   $N=15$  case we found no singular solutions.  Our approach can also  be tested for higher values  of the length of the spin chain. 

%------------------------------------------------------------------------------------------------------------------------------------------------------------------------------------------

\section{Acknowledgement} 
The present study is partially supported by Grant-in-Aid for Scientific Research No. 24540396.
P. Giri acknowledges the financial support from JSPS. We would like to thank the anonymous referee for pointing out   mistakes/typos   in some equations and for valuable suggestions.

%------------------------------------------------------------------------------------------------------------------------------------------------------------------------------------------
\appendix 

%------------------------------------------------------------------------------------------------------------------------------------------------------------------------------------------

\section{Two down-spin singular state for even N}\label{app1}
%\section{Appendix}\label{app1}
%\subsection{Two down-spin singular state for even N}
Here,  up to a proportionality constant,  we show   eq. (\ref{sgnvec1})   with the help of  eq.  (\ref{vec1}),   (\ref{vec}) and (\ref{eregunepo}). Let us start with the definition of the singular Bethe state    (\ref{vec}) for two down spins, 
\begin{eqnarray}\label{apvec1}
|\frac{i}{2},-\frac{i}{2}\rangle &=& \lim_{\epsilon\to 0}\frac{1}{(\tilde{\lambda}_1-\frac{1}{2}i)^N}B(\tilde{\lambda}_1)B(\tilde{\lambda}_2)|\Omega\rangle \,.
 \end{eqnarray}
Substituting  explicit expression for  the two down spin Bethe  eigenstate,  obtained from  (\ref{vec1}),  in the above equation, we obtain
  \begin{eqnarray}\nonumber
|\frac{i}{2},-\frac{i}{2}\rangle &=& - \lim_{\epsilon\to 0} \frac{1}{(\tilde{\lambda}_1-\frac{1}{2}i)^N}\frac{\tilde{\lambda}_1-\tilde{\lambda}_2 +i}{\tilde{\lambda}_1-\tilde{\lambda}_2} \frac{(\tilde{\lambda}_1- \frac{i}{2})^N}{\tilde{\lambda}_1+ \frac{i}{2}}\frac{(\tilde{\lambda}_2-\frac{i}{2})^N}{\tilde{\lambda}_2+\frac{i}{2}}
 \sum_{1\leq x_1 < x_2\leq N}^{N}  \left[ \left(\frac{\tilde{\lambda}_{1}+ \frac{i}{2}}{\tilde{\lambda}_{1}-\frac{i}{2}}\right)^{x_1}\left(\frac{\tilde{\lambda}_{2}+\frac{i}{2}}{\tilde{\lambda}_{2}-\frac{i}{2}}\right)^{x_2} \right. \\ \label{apvec2}
 && \left. +\frac{\tilde{\lambda}_{1}-\tilde{\lambda}_{2}-i}{\tilde{\lambda}_{1}-\tilde{\lambda}_{2}+i}\left(\frac{\tilde{\lambda}_{2}+\frac{i}{2}}{\tilde{\lambda}_{2}-\frac{i}{2}}\right)^{x_1}\left(\frac{\tilde{\lambda}_{1}+\frac{i}{2}}{\tilde{\lambda}_{1}-\frac{i}{2}}\right)^{x_2}  \right]\prod_{j=1}^2 S^-_{x_j}|\Omega \rangle\,.
 \end{eqnarray}
Replacing   $\tilde{\lambda}_1, \tilde{\lambda}_2$ of   eq. (\ref{eregunepo})  in (\ref{apvec2}) we obtain 
 \begin{eqnarray} \nonumber
|\frac{i}{2},-\frac{i}{2}\rangle &=&   2 (-1)^{N/2}\lim_{\epsilon\to 0}  \sum_{1\leq x_1 < x_2\leq N}^{N}  \left[ (-1)^{x_2} \epsilon^{x_2-x_1-1} \left(1+ o(\epsilon) + h.o (\epsilon)\right) \right.\\  \label{apvec3}
 &&\left. + (-1)^{x_1} \epsilon^{N+x_1-x_2-1} \left(1+ o(\epsilon) + h.o (\epsilon)\right) \right]\prod_{j=1}^2 S^-_{x_j}|\Omega \rangle\,,
 \end{eqnarray} 
where $o(\epsilon)$ and $h.o(\epsilon)$ are the order $\epsilon$ and higher oder terms respectively.  Taking the $\epsilon  \to 0$ limit  in (\ref{apvec3})  we observe  that    the first term survives for   $x_2=x_1+1$   and  in second  term survives for  $x_1=1, x_2=N$. Finally  we obtain
 \begin{eqnarray} \label{apvec4}
|\frac{i}{2},-\frac{i}{2}\rangle = 2 (-1)^{N/2+1} \sum_{j}^N \left(-1\right)^jS^-_{j} S^-_{j+1}|\Omega \rangle\,.
 \end{eqnarray}
%---------------------------------------------------------------------------------------------------------------------------------------------------------------------------------------------

\section{Three down-spin  singular state for even  N}\label{app2}
We now prove, up to a proportionality constant,    eq. (\ref{sgnvec12}). Let us start with the definition of the singular Bethe  eigenstate    (\ref{vec}) for $N$ even and  three down spins
\begin{eqnarray}\label{apvec12si}
|\frac{i}{2},-\frac{i}{2},0\rangle &=& \lim_{\epsilon\to 0}\frac{1}{(\tilde{\lambda}_1-\frac{1}{2}i)^N}B(\tilde{\lambda}_1)B(\tilde{\lambda}_2)B(0)|\Omega\rangle\,.
\end{eqnarray}
Substituting  explicit form of eq.  (\ref{vec1}) and because the third  rapidity of a singular solution for  an even length chain vanishes,  setting    $\lambda_3=0$ in the above equation we obtain
\begin{eqnarray}\label{apvec12} \nonumber
|\frac{i}{2},-\frac{i}{2},0\rangle &=&  i \lim_{\epsilon\to 0} \frac{1}{(\tilde{\lambda}_1-\frac{1}{2}i)^N}\frac{\tilde{\lambda}_1-\tilde{\lambda}_2 +i}{\tilde{\lambda}_1-\tilde{\lambda}_2} \frac{\tilde{\lambda}_1 +i}{\tilde{\lambda}_1}
\frac{\tilde{\lambda}_2 +i}{\tilde{\lambda}_2}
\frac{(\tilde{\lambda}_1-\frac{i}{2})^N}{\tilde{\lambda}_1+\frac{i}{2}}\frac{(\tilde{\lambda}_2-\frac{i}{2})^N}{\tilde{\lambda}_2+\frac{i}{2}}\frac{(-\frac{i}{2})^N}{\frac{i}{2}}
 \sum_{1\leq x_1 < x_2 <x_3\leq N}^{N} \\ \nonumber
  &&\left[ \left(\frac{\tilde{\lambda}_{1}+ \frac{i}{2}}{\tilde{\lambda}_{1}-\frac{i}{2}}\right)^{x_1}\left(\frac{\tilde{\lambda}_{2}+\frac{i}{2}}{\tilde{\lambda}_{2}-\frac{i}{2}}\right)^{x_2}(-1)^{x_3} + 
 \frac{\tilde{\lambda}_{1}-\tilde{\lambda}_{2}-i}{\tilde{\lambda}_{1}-\tilde{\lambda}_{2}+i}  \frac{\tilde{\lambda}_{1}-i}{\tilde{\lambda}_{1}+i} 
\left(\frac{\tilde{\lambda}_{2}+\frac{i}{2}}{\tilde{\lambda}_{2}-\frac{i}{2}}\right)^{x_1} \left(-1\right)^{x_2}\left(\frac{\tilde{\lambda}_{1}+\frac{i}{2}}{\tilde{\lambda}_{1}-\frac{i}{2}}\right)^{x_3}  + \right.\\ \nonumber
&& \left.  \frac{\tilde{\lambda}_{1}-i}{\tilde{\lambda}_{1}+i} \frac{\tilde{\lambda}_{2}-i}{\tilde{\lambda}_{2}+i} (-1)^{x_1}
\left(\frac{\tilde{\lambda}_{1}+\frac{i}{2}}{\tilde{\lambda}_{1}-\frac{i}{2}}\right)^{x_2} \left(\frac{\tilde{\lambda}_{2}+\frac{i}{2}}{\tilde{\lambda}_{2}-\frac{i}{2}}\right)^{x_3}  +
   \frac{\tilde{\lambda}_{2}-i}{\tilde{\lambda}_{2}+i} 
\left(\frac{\tilde{\lambda}_{1}+\frac{i}{2}}{\tilde{\lambda}_{1}-\frac{i}{2}}\right)^{x_1} (-1)^{x_2} \left(\frac{\tilde{\lambda}_{2}+\frac{i}{2}}{\tilde{\lambda}_{2}-\frac{i}{2}}\right)^{x_3}  + \right.\\  \nonumber
&& \left. \frac{\tilde{\lambda}_{1}-\tilde{\lambda}_{2}-i}{\tilde{\lambda}_{1}-\tilde{\lambda}_{2}+i}  \left(\frac{\tilde{\lambda}_{2}+\frac{i}{2}}{\tilde{\lambda}_{2}-\frac{i}{2}}\right)^{x_1}\left(\frac{\tilde{\lambda}_{1}+\frac{i}{2}}{\tilde{\lambda}_{1}-\frac{i}{2}}\right)^{x_2}(-1)^{x_3}+  \right.\\ 
&&\left.  \frac{\tilde{\lambda}_{1}-\tilde{\lambda}_{2}-i}{\tilde{\lambda}_{1}-\tilde{\lambda}_{2}+i}  \frac{\tilde{\lambda}_{1}-i}{\tilde{\lambda}_{1}+i} 
 \frac{\tilde{\lambda}_{2}-i}{\tilde{\lambda}_{2}+i} (-1)^{x_1}
 \left(\frac{\tilde{\lambda}_{2}+\frac{i}{2}}{\tilde{\lambda}_{2}-\frac{i}{2}}\right)^{x_2}\left(\frac{\tilde{\lambda}_{1}+\frac{i}{2}}{\tilde{\lambda}_{1}-\frac{i}{2}}\right)^{x_3} 
\right]\prod_{j=1}^3 S^-_{x_j}|\Omega \rangle\,.
\end{eqnarray}
Replacing  the explicit form  (\ref{regunepo3}) in eq. (\ref{apvec12}) and expanding  in powers  $\epsilon$    we obtain
\begin{eqnarray}\label{apvec13} \nonumber
|\frac{i}{2},-\frac{i}{2},0\rangle &=&  3\times  2^{2-N}i^{N-1}\lim_{\epsilon\to 0}  \sum_{1\leq x_1 < x_2 <x_3\leq N}^{N}  \left[ \left(\sqrt[N]{-\frac{1}{3}} \epsilon\right)^{x_2-x_1-1} (-1)^{x_2+x_3}  \left(1 + h.o \right) + \right.\\  \nonumber
&&\left.   \left(\sqrt[N]{-\frac{1}{3}} \epsilon \right)^{N+ x_1-x_3-1} (-1)^{x_1+x_2}  \left(1+  h.o \right) + 
   \left(\sqrt[N]{-\frac{1}{3}} \epsilon \right)^{x_3-x_2-1} (-1)^{x_1-x_3} \left(1+h.o \right) + \right.\\  \nonumber
&&\left.   3 \left(\sqrt[N]{-\frac{1}{3}} \epsilon \right)^{x_3-x_1-1} (-1)^{x_2+x_3+1}  \left(1+ h.o \right) +
   \left(\sqrt[N]{-\frac{1}{3}} \epsilon \right)^{x_1-x_2-1} (-1)^{x_3+x_1} \epsilon^{N} \left(1+h.o \right) + \right.\\  
&&\left.    \left(\sqrt[N]{-\frac{1}{3}}\epsilon \right)^{x_2-x_3-1} (-1)^{x_1+x_2+1} \epsilon^{N} \left(1+ h.o \right)  \right]\prod_{j=1}^3 S^-_{x_j}|\Omega \rangle\,, 
\end{eqnarray}
where   $h.o$ represents  terms of order   $o(\epsilon) + h.o(\epsilon)$.  Taking the  $\epsilon\to 0$  limit  in (\ref{apvec13}) we see that the first term survives for $x_2=x_1+1$,  the second term survives for $x_1=1, x_3=N$,  the  third term survives 
for $x_3= x_2+1$ and the remaining last three terms vanish.  We therefore  obtain
\begin{eqnarray}\label{apvec14} \nonumber
|\frac{i}{2},-\frac{i}{2},0\rangle &=&  3\times  2^{2-N}i^{N+1}  \left[ \sum_{1\leq x_1 < x_1+1 <x_3\leq N}^{N}    (-1)^{x_1+x_3} S^-_{x_1}S^-_{x_1+1}S^-_{x_3}   + \right. \\ 
&& \left. \sum_{2\leq x_2 \leq N-1}^{N}    (-1)^{x_2} S^-_{1}S^-_{x_2}S^-_{N}  +
\sum_{1\leq x_1 < x_2 <x_2+1\leq N}^{N}    (-1)^{x_1+x_2} S^-_{x_1}S^-_{x_2}S^-_{x_2+1}  \right] |\Omega \rangle\,.
\end{eqnarray}
Finally we obtain the simplified form of the singular state for three down spins 
\begin{eqnarray}\label{apvec15} 
|\frac{i}{2},-\frac{i}{2},0\rangle &=&  3\times  2^{2-N}i^{N+1}  \left(\sum_{k=1}^N (-1)^k S^-_k\right) \sum_{j=1}^N (-1)^j S^-_jS^-_{j+1}|\Omega\rangle\,.
\end{eqnarray}

%---------------------------------------------------------------------------------------------------------------------------------------------------------------------------------------------

%---------------------------------------------------------------------------------------------------------------------------------------------------------------------------------------------

\section{Condition for the three down-spin singular states for odd N}\label{app3}
In this appendix we  prove  eq. (\ref{oddfac1})  and its  corresponding condition  eq. (\ref{oddfac2}). Let is start with    eq. (\ref{bethe2})
\begin{eqnarray}\label{ap1bethe2}\nonumber
0 &=&\left(\lambda_3- \frac{1}{2}i\right)^N \left(\lambda_3+ \frac{1}{2}i\right)\left(\lambda_3+\frac{3}{2}i\right)  -\left(\lambda_3 + \frac{1}{2}i\right)^N \left(\lambda_3 - \frac{1}{2}i\right)\left(\lambda_3-\frac{3}{2}i\right)\\
\nonumber
&=& \left(\lambda_3- \frac{1}{2}i\right)^N \left(\lambda_3^2 +2i\lambda_3-\frac{3}{4}\right)- \left(\lambda_3+ \frac{1}{2}i\right)^N \left(\lambda_3^2 -2i\lambda_3-\frac{3}{4}\right)\\
&=& \left(\lambda_3^2-\frac{3}{4}\right) \left[\left(\lambda_3- \frac{1}{2}i\right)^N - \left(\lambda_3+ \frac{1}{2}i\right)^N\right] + 2i\lambda_3 \left[\left(\lambda_3- \frac{1}{2}i\right)^N + \left(\lambda_3+ \frac{1}{2}i\right)^N\right] \,.
\end{eqnarray}
The first term of  eq. (\ref{ap1bethe2})  already  has  the desired factor   $\left(\lambda_3^2-3/4\right)$.  To find   out the same factor in the second term  let us consider 
\begin{eqnarray}\label{ap2bethe2}\nonumber
\left(\lambda_3- \frac{1}{2}i\right)^N + \left(\lambda_3+ \frac{1}{2}i\right)^N  &=& \sum_{p=0}^{N}\Comb{N}{p} {\lambda_3}^{N-p} \left(\frac{i}{2}\right)^p \left[ (-1)^p + 1\right]
= 2\sum_{r=0}^{\frac{N-1}{2}}\Comb{N}{2r} {\lambda_3}^{N-2r} \left(\frac{i}{2}\right)^{2r}\,,  ~~~~~p= 2r\\ \nonumber
&=& 2\lambda_3\sum_{r=0}^{\frac{N-1}{2}} \frac{1}{4^r}\Comb{N}{2r}  \left(-1\right)^r{\lambda_3}^{N-1-2r}\,, \\ \nonumber
&=&  2\lambda_3\left[\frac{1}{4^0}\Comb{N}{0}  \left(-1\right)^0{\lambda_3}^{N-1} + \frac{1}{4^1}\Comb{N}{2}  \left(-1\right)^1{\lambda_3}^{N-3} + \cdots + \frac{1}{4^{\frac{N-1}{2}}}\Comb{N}{N-1}  \left(-1\right)^{\frac{N-1}{2}} \right]\\ \nonumber
&=&  2\lambda_3\left[\frac{1}{4^0}\Comb{N}{0}  \left(-1\right)^0{\lambda_3}^{N-3} \left(\lambda_3^2-\frac{3}{4}\right) +  \right. \\ \nonumber
&&\left. \left(\frac{1}{4^1}\Comb{N}{2}  \left(-1\right)^1 + 
\frac{3}{4}\frac{1}{4^0}\Comb{N}{0}  \left(-1\right)^0\right){\lambda_3}^{N-5}\left(\lambda_3^2-\frac{3}{4}\right)  + \cdots + \right. \\ \nonumber
&&\left. \left(\frac{1}{4^{\frac{N-3}{2}}} \sum_{s=0}^{\frac{N-3}{2}}\Comb{N}{2s}  \left(-1\right)^s 3^{\frac{N-3}{2}-s}  \right){\lambda_3}^{N-5}\left(\lambda_3^2-\frac{3}{4}\right)\right]\\ 
&=& 2\lambda_3 \left(\lambda_3^2-\frac{3}{4}\right) \left[\sum_{r=0}^{\frac{N-3}{2}}
\lambda_3^{N-3-2r}\left(\frac{3}{4}\right)^r\left(\sum_{s=0}^{r}\left(-\frac{1}{3}\right)^s \Comb{N}{2s}\right)\right]\,,
\end{eqnarray}
where $\Comb{x}{y}$ is the binomial coefficient.  Substituting  the last expression of  (\ref{ap2bethe2})  back in  eq.  (\ref{ap1bethe2})   we obtain    eq.  (\ref{oddfac1}).  In order to arrive at  the last expression of  (\ref{ap2bethe2})  we  need a matching condition  at the end of the series  expansion, which is  given by 
\begin{eqnarray}\label{ap3bethe2}\nonumber
0 &=&  -3^{\frac{N-1}{2}}\sum_{s=0}^{\frac{N-3}{2}}\left(-\frac{1}{3}\right)^{s} \Comb{N}{2s} -N (-1)^{\frac{N-1}{2}}\,,\\ 
&=& -3^{\frac{N-1}{2}}\sum_{s=0}^{\frac{N-1}{2}}\left(-\frac{1}{3}\right)^{s} \Comb{N}{2s}\,, \\ \label{ap4bethe2}
&=& -3^{\frac{N-1}{2}}  {_2 F_1} \left(-\frac{N}{2}, \frac{1-N}{2}, \frac{1}{2}, -\frac{1}{3}\right)\,, \\  \label{ap5bethe2}
&=&  -\frac{2^N}{\sqrt{3}} \cos \left(\frac{\pi}{6}N\right)\,.
\end{eqnarray} 
To arrive at  expression  (\ref{ap4bethe2})  from  (\ref{ap3bethe2})   we have used the relation   {\bf 15.4.1} of \cite{abra}. Note that   $a$ or $b$ of   ${_2F_1(a,b,c,z)}$  has to be negative in order to hold the relation. 
In our case  since $N \geq 9$ is odd,    $b= (1-N)/2$ is always a  negative integer.   To obtain  (\ref{ap5bethe2})  from (\ref{ap4bethe2})  we have used the relation    {\bf 15.1.19}  of  \cite{abra}.  Eq. (\ref{ap5bethe2}) is satisfied when 
the length, $N$,  of the spin chain  is given by    eq. (\ref{oddfac2}).

%------------------------------------------------------------------------------------------------------------------------------------------------------------------------------------------

\section{Three down-spin  singular states for odd  N}\label{app21}
We now prove, up to a proportionality constant,    eq. (\ref{sgnveco12}). Let us start with the definition of the singular Bethe  eigenstate    (\ref{vec}) for odd-$N$  and  three down spins
\begin{eqnarray}\label{apvec12sio} 
|\frac{i}{2},-\frac{i}{2}, \pm \frac{\sqrt{3}}{2}\rangle &=& \lim_{\epsilon\to 0}\frac{1}{(\tilde{\lambda}_1-\frac{1}{2}i)^N}B(\tilde{\lambda}_1)B(\tilde{\lambda}_2)B(\pm \frac{\sqrt{3}}{2})|\Omega\rangle\,.
\end{eqnarray}
Substituting  explicit form of eq.  (\ref{vec1}) and  setting    $\lambda_3= \pm \frac{\sqrt{3}}{2}$ in the above equation we obtain
\begin{eqnarray}\label{apvec12o} \nonumber
&& |\frac{i}{2},-\frac{i}{2}, \pm \frac{\sqrt{3}}{2}\rangle = \\ \nonumber
&& i \lim_{\epsilon\to 0} \frac{1}{(\tilde{\lambda}_1-\frac{1}{2}i)^N}\frac{\tilde{\lambda}_1-\tilde{\lambda}_2 +i}{\tilde{\lambda}_1-\tilde{\lambda}_2} \frac{\tilde{\lambda}_1 \mp \frac{\sqrt{3}}{2} +i}{\tilde{\lambda}_1 \mp \frac{\sqrt{3}}{2}}
\frac{\tilde{\lambda}_2 \mp \frac{\sqrt{3}}{2} +i}{\tilde{\lambda}_2 \mp \frac{\sqrt{3}}{2}}
\frac{(\tilde{\lambda}_1-\frac{i}{2})^N}{\tilde{\lambda}_1+\frac{i}{2}}\frac{(\tilde{\lambda}_2-\frac{i}{2})^N}{\tilde{\lambda}_2+\frac{i}{2}}\frac{( \pm \frac{\sqrt{3}}{2}-\frac{i}{2})^N}{\pm \frac{\sqrt{3}}{2} +\frac{i}{2}}
 \sum_{1\leq x_1 < x_2 <x_3\leq N}^{N} \\ \nonumber
  &&\left[ \left(\frac{\tilde{\lambda}_{1}+\frac{i}{2}}{\tilde{\lambda}_{1}-\frac{i}{2}}\right)^{x_1}\left(\frac{\tilde{\lambda}_{2}+\frac{i}{2}}{\tilde{\lambda}_{2}-\frac{i}{2}}\right)^{x_2}(\exp{(\pm \frac{\pi}{3}i)})^{x_3} + 
 \frac{\tilde{\lambda}_{1}-\tilde{\lambda}_{2}-i}{\tilde{\lambda}_{1}-\tilde{\lambda}_{2}+i}  \frac{\tilde{\lambda}_{1} \mp \frac{\sqrt{3}}{2}-i}{\tilde{\lambda}_{1} \mp \frac{\sqrt{3}}{2}+i} 
\left(\frac{\tilde{\lambda}_{2}+\frac{i}{2}}{\tilde{\lambda}_{2}-\frac{i}{2}}\right)^{x_1} \left( \exp{(\pm \frac{\pi}{3}i)}\right)^{x_2}\left(\frac{\tilde{\lambda}_{1}+\frac{i}{2}}{\tilde{\lambda}_{1}-\frac{i}{2}}\right)^{x_3}  + \right.\\ \nonumber
&& \left.  \frac{\tilde{\lambda}_{1} \mp \frac{\sqrt{3}}{2}-i}{\tilde{\lambda}_{1} \mp \frac{\sqrt{3}}{2}+i} \frac{\tilde{\lambda}_{2} \mp \frac{\sqrt{3}}{2}-i}{\tilde{\lambda}_{2} \mp \frac{\sqrt{3}}{2}+i} (\exp{(\pm \frac{\pi}{3}i)})^{x_1}
\left(\frac{\tilde{\lambda}_{1}+ \frac{i}{2}}{\tilde{\lambda}_{1}-\frac{i}{2}}\right)^{x_2} \left(\frac{\tilde{\lambda}_{2}+\frac{i}{2}}{\tilde{\lambda}_{2}-\frac{i}{2}}\right)^{x_3}  + \right.\\ \nonumber
&& \left.   \frac{\tilde{\lambda}_{2} \mp \frac{\sqrt{3}}{2}-i}{\tilde{\lambda}_{2} \mp \frac{\sqrt{3}}{2}+i} \left(\frac{\tilde{\lambda}_{1} 
+\frac{i}{2}}{\tilde{\lambda}_{1}-\frac{i}{2}}\right)^{x_1} (\exp{(\pm \frac{\pi}{3}i)})^{x_2} \left(\frac{\tilde{\lambda}_{2}+\frac{i}{2}}{\tilde{\lambda}_{2}-\frac{i}{2}}\right)^{x_3}  + 
 \frac{\tilde{\lambda}_{1}-\tilde{\lambda}_{2}-i}{\tilde{\lambda}_{1}-\tilde{\lambda}_{2}+i}  \left(\frac{\tilde{\lambda}_{2}+\frac{i}{2}}{\tilde{\lambda}_{2}-\frac{i}{2}}\right)^{x_1}\left(\frac{\tilde{\lambda}_{1}+ \frac{i}{2}}{\tilde{\lambda}_{1}-\frac{i}{2}}\right)^{x_2}( \exp{(\pm \frac{\pi}{3}i)})^{x_3}+  \right.\\ 
&&\left.  \frac{\tilde{\lambda}_{1}-\tilde{\lambda}_{2}-i}{\tilde{\lambda}_{1}-\tilde{\lambda}_{2}+i}  \frac{\tilde{\lambda}_{1} \mp \frac{\sqrt{3}}{2}-i}{\tilde{\lambda}_{1} \mp \frac{\sqrt{3}}{2}+i} 
 \frac{\tilde{\lambda}_{2} \mp \frac{\sqrt{3}}{2}-i}{\tilde{\lambda}_{2} \mp \frac{\sqrt{3}}{2}+i} (\exp{(\pm \frac{\pi}{3}i)})^{x_1}
 \left(\frac{\tilde{\lambda}_{2}+\frac{i}{2}}{\tilde{\lambda}_{2}-\frac{i}{2}}\right)^{x_2}\left(\frac{\tilde{\lambda}_{1}+\frac{i}{2}}{\tilde{\lambda}_{1}-\frac{i}{2}}\right)^{x_3} 
\right]\prod_{j=1}^3 S^-_{x_j}|\Omega \rangle\,.
\end{eqnarray}
Replacing  the explicit form  (\ref{regunepo}) in eq. (\ref{apvec12o}) and expanding  in powers  of $\epsilon$    we obtain
\begin{eqnarray}\label{apvec13o} \nonumber
|\frac{i}{2},-\frac{i}{2}, \pm \frac{\sqrt{3}}{2}\rangle &=&  \mp  2\sqrt{3} (-1)^{\frac{N+1}{2}}\exp{\left(\mp\frac{\pi}{6}(N+1)\right)} \times \\ \nonumber
&&\lim_{\epsilon\to 0}  \sum_{1\leq x_1 < x_2 <x_3\leq N}^{N}  \left[ \left(i\sqrt[N]{\pm \frac{1}{\sqrt{3}}} \epsilon\right)^{x_2-x_1-1} \left(\exp{(\pm \frac{\pi}{3}i)}\right)^{x_3} i^{x_1+x_2-1}  \left(1 + h.o \right) + \right.\\  \nonumber
&&\left.   \left(i\sqrt[N]{ \pm \frac{1}{\sqrt{3}}} \epsilon \right)^{N+ x_1-x_3-1}  \left(\exp{(\pm \frac{\pi}{3}i)}\right)^{x_2} i^{x_1+x_3-N-1}  \left(1+  h.o \right) +\right. \\ \nonumber
  &&\left. \left(i\sqrt[N]{\pm\frac{1}{\sqrt{3}}} \epsilon \right)^{x_3-x_2-1} \left(\exp{(\pm \frac{\pi}{3}i)}\right)^{x_1}i^{-2N+x_2+x_3-1} \left(1+h.o \right) + \right.\\  \nonumber
&&\left.    \left(i\sqrt[N]{\pm\frac{1}{\sqrt{3}}} \epsilon \right)^{x_3-x_1-1}\left(\exp{(\pm \frac{\pi}{3}i)}\right)^{x_2}i^{x_1+x_3-1}  \left(1+ h.o \right) + \right.\\ \nonumber
&& \left.   \left(i\sqrt[N]{\pm\frac{1}{\sqrt{3}}} \epsilon \right)^{x_1-x_2-1} \left(\exp{(\pm \frac{\pi}{3}i)}\right)^{x_3}i^{x_1+x_2-1} \epsilon^{N} \left(1+h.o \right) + \right.\\  
&&\left.    \left(i\sqrt[N]{\pm\frac{1}{\sqrt{3}}}\epsilon \right)^{x_2-x_3-1} \left(\exp{(\pm \frac{\pi}{3}i)}\right)^{x_1}
i^{x_2+x_3+1} \epsilon^{N} \left(1+ h.o \right)  \right]\prod_{j=1}^3 S^-_{x_j}|\Omega \rangle\,, 
\end{eqnarray}
where   $h.o$ represents  terms of order   $o(\epsilon) + h.o(\epsilon)$.  Taking the  $\epsilon\to 0$  limit  in (\ref{apvec13o}) we see that the first term survives for $x_2=x_1+1$,  the second term survives for $x_1=1, x_3=N$,  the  third term survives 
for $x_3= x_2+1$ and the remaining last three terms vanish.  We therefore  obtain
\begin{eqnarray}\label{apvec14o} \nonumber
&&|\frac{i}{2},-\frac{i}{2}, \frac{\sqrt{3}}{2}\rangle =  \mp  2\sqrt{3} (-1)^{\frac{N+1}{2}}\exp{\left(\mp\frac{\pi}{6}(N+1)\right)}
 \left[ \sum_{1\leq x_1 < x_1+1 <x_3\leq N}^{N}    (-1)^{x_1} \left(\exp{(\pm \frac{\pi}{3}i)}\right)^{x_3} S^-_{x_1}S^-_{x_1+1}S^-_{x_3}   + \right. \\ 
&& \left. \sum_{2\leq x_2 \leq N-1}^{N}  \left(\exp{(\pm \frac{\pi}{3}i)}\right)^{x_2}  S^-_{1}S^-_{x_2}S^-_{N}  +
\sum_{1\leq x_1 < x_2 <x_2+1\leq N}^{N}    (-1)^{x_2+1} \left(\exp{(\pm \frac{\pi}{3}i)}\right)^{x_1}S^-_{x_1}S^-_{x_2}S^-_{x_2+1}  \right] |\Omega \rangle\,.
\end{eqnarray}
Finally we obtain the simplified form of  (\ref{apvec14o})  for the singular states of  three down spins 
\begin{eqnarray}\label{apvec15o} 
|\frac{i}{2},-\frac{i}{2},\pm \frac{\sqrt{3}}{2}\rangle =\mp  2\sqrt{3} (-1)^{\frac{N+1}{2}}\exp{\left(\mp\frac{\pi}{6}(N+1)\right)}\sum_{k=1}^N \left( \exp{(\pm \frac{\pi}{3}ki)}S^-_k\sum_{j=1}^N (-1)^{j+H(j-k)} S^-_jS^-_{j+1}\right)|\Omega\rangle\,.
\end{eqnarray}

%---------------------------------------------------------------------------------------------------------------------------------------------------------------------------------------------

\end{document}